# Reasoning About Knowledge of Unawareness[*]

Joseph Y. Halpern
Computer Science Department
Cornell University, U.S.A.
e-mail: halpern@cs.cornell.edu

Leandro Chaves Rêgo
School of Electrical and Computer Engineering
Cornell University, U.S.A.
e-mail: lcr26@cornell.edu

September 24, 2018

**Abstract**

Awareness has been shown to be a useful addition to standard epistemic logic for many applications. However, standard propositional logics for knowledge and awareness cannot express the fact that an agent knows that there are facts of which he is unaware without there being an explicit fact that the agent knows he is unaware of. We propose a logic for reasoning about knowledge of unawareness, by extending Fagin and Halpern's *Logic of General Awareness*. The logic allows quantification over variables, so that there is a formula in the language that can express the fact that "an agent explicitly knows that there exists a fact of which he is unaware". Moreover, that formula can be true without the agent explicitly knowing that he is unaware of any particular formula. We provide a sound and complete axiomatization of the logic, using standard axioms from the literature to capture the quantification operator. Finally, we show that the validity problem for the logic is recursively enumerable, but not decidable.

[*]This work was supported in part by NSF under grants CTC-0208535, ITR-0325453, and IIS-0534064, by ONR under grants N00014-00-1-03-41 and N00014-01-10-511, and by the DoD Multidisciplinary University Research Initiative (MURI) program administered by the ONR under grant N00014-01-1-0795. The second author was also supported in part by a scholarship from the Brazilian Government through the Conselho Nacional de Desenvolvimento Científico e Tecnológico (CNPq).

# 1 Introduction

As is well known, standard models of epistemic logic suffer from the *logical omniscience* problem (first observed and named by Hintikka [1962]): agents know all tautologies and all the logical consequences of their knowledge. This seems inappropriate for resource-bounded agents and agents who are unaware of various concepts (and thus do not know logical tautologies involving those concepts). Many approaches to avoiding this problem have been suggested. One of the best-known is due to Fagin and Halpern [1988] (FH from now on). It involves distinguishing *explicit knowledge* from *implicit knowledge*, using a syntactic awareness operator. Roughly speaking, implicit knowledge is the standard (S5) notion of knowledge; explicit knowledge amounts to implicit knowledge and *awareness*.

Since this approach was first introduced by FH, there has been a stream of papers on awareness in the economics literature (see, for example, [Dekel, Lipman, and Rustichini 1998; Halpern 2001; Heifetz, Meier, and Schipper 2003; Modica and Rustichini 1994; Modica and Rustichini 1999]). The logics used in these papers cannot express the fact that an agent may (explicitly) know that he is unaware of some facts. Indeed, in the language of [Halpern 2001], all of these models are special cases of the original awareness model where awareness is *generated by primitive propositions*, that is, an agent is aware of a formula iff the agent is aware of all primitive propositions that appear in the formula. If awareness is generated by primitive propositions, then it is impossible for an agent to know that he is unaware of a specific fact.

Nevertheless, knowledge of unawareness comes up often in real-life situations. For example, when a primary physician sends a patient to an expert on oncology, he knows that an oncologist is aware of things that could help the patient's treatment of which he is not aware. Moreover, the physician is unlikely to know which specific thing he is unaware of that would improve the patient's condition (if he knew which one it was, he would not be unaware of it!). Similarly, when an investor decides to let his money be managed by an investment fund company, he knows the company is aware of more issues involving the financial market than he is (and is thus likely to get better results with his money), but again, the investor is unlikely to be aware of the specific relevant issues. Ghirardato [2001] pointed out the importance of dealing with unawareness and knowledge of unawareness in the context of decision-making, but did not give a formal model.

To model knowledge of unawareness, we extend the syntax of the logic of general awareness considered by FH to allow for quantification over variables. Thus, we allow formulas such as $X_i(\exists x \neg A_i x)$, which says that agent $i$ (explicitly) knows that there exists a formula of which he is not aware. The idea of adding propositional quantification to modal logic is well known in the literature (see, for example, [Bull 1969; Engelhardt, Meyden, and Moses 1998; Fine 1970; Kaplan 1970; Kripke 1959]). However, as we explain in Section 3, because $A_i$ is a syntactic operator, we are forced to give somewhat nonstandard semantics to the existential operator. Nevertheless, we are able to provide a sound and complete axiomatization of the resulting logic, using standard axioms from the literature to capture the quantification operator. Using the logic, we can easily characterize the knowledge of the relevant agents in all the examples we consider.

The rest of the paper is organized as follows. In Section 2, we review the standard semantics for knowledge and awareness. In Section 3, we introduce our logic for reasoning about knowledge of unawareness. In Section 4 we axiomatize the logic, and in Section 5, we consider the complexity of the decision problem for the logic. We conclude in Section 6.

## 2 The FH model

We briefly review the FH Logic of General Awareness here, before generalizing it to allow quantification over propositional variables. The syntax of the logic is as follows: given a set $\{1, \ldots, n\}$ of agents, formulas are formed by starting with a set $\Phi = \{p, q, \ldots\}$ of primitive propositions, and then closing off under conjunction ($\wedge$), negation ($\neg$), and the modal operators $K_i, A_i, X_i, i = 1, \ldots, n$. Call the resulting language $\mathcal{L}_n^{K,X,A}(\Phi)$. As usual, we define $\varphi \vee \psi$ and $\varphi \Rightarrow \psi$ as abbreviations of $\neg(\neg\varphi \wedge \neg\psi)$ and $\neg\varphi \vee \psi$, respectively. The intended interpretation of $A_i\varphi$ is "$i$ is aware of $\varphi$". The power of this approach comes from the flexibility of the notion of awareness. For example, "agent $i$ is aware of $\varphi$" may be interpreted as "agent $i$ is familiar with all primitive propositions in $\varphi$" or as "agent $i$ can compute the truth value of $\varphi$ in time $t$".

Having awareness in the language allows us to distinguish two notions of knowledge: implicit knowledge and explicit knowledge. Implicit knowledge, denoted by $K_i$, is defined as truth in all states that the agent considers possible, as usual. Explicit knowledge, denoted by $X_i$, is defined as the conjunction of implicit knowledge and awareness.

We give semantics to formulas in $\mathcal{L}_n^{K,X,A}(\Phi)$ in awareness structures. A tuple $M = (S, \pi, \mathcal{K}_1, ..., \mathcal{K}_n, \mathcal{A}_1, \ldots, \mathcal{A}_n)$ is an *awareness structure for $n$ agents (over $\Phi$)* if $S$ is a set of states, $\pi : S \times \Phi \to \{\text{true}, \text{false}\}$ is an interpretation that determines which primitive propositions are true at each state, $\mathcal{K}_i$ is a binary relation on $S$ for each agent $i = 1, \ldots, n$, and $\mathcal{A}_i$ is a function associating a set of formulas with each state in $S$, for $i = 1, ..., n$. Intuitively, if $(s, t) \in \mathcal{K}_i$, then agent $i$ considers state $t$ possible at state $s$, while $\mathcal{A}_i(s)$ is the set of formulas that agent $i$ is aware of at state $s$. The set of formulas the agent is aware of can be arbitrary. Depending on the interpretation of awareness one has in mind, certain restrictions on $\mathcal{A}_i$ may apply. (We discuss some interesting restrictions in the next section.)

Let $\mathcal{M}_n(\Phi)$ denote the class of all awareness structures for $n$ agents over $\Phi$, with no restrictions on the $\mathcal{K}_i$ relations and on the functions $\mathcal{A}_i$. We use the superscripts $r$, $e$, and $t$ to indicate that the $\mathcal{K}_i$ relations are restricted to being reflexive, Euclidean,[1] and transitive, respectively. Thus, for example, $\mathcal{M}_n^{rt}(\Phi)$ is the class of all reflexive and transitive awareness structures for $n$ agents.

We write $(M, s) \models \varphi$ if $\varphi$ is true at state $s$ in the awareness structure $M$. The truth relation is defined inductively as follows:

$$(M, s) \models p, \text{ for } p \in \Phi, \text{ if } \pi(s, p) = \text{true}$$
$$(M, s) \models \neg\varphi \text{ if } (M, s) \not\models \varphi$$
$$(M, s) \models \varphi \wedge \psi \text{ if } (M, s) \models \varphi \text{ and } (M, s) \models \psi$$

---
[1] Recall that a binary relation $\mathcal{K}_i$ is Euclidean if $(s, t), (s, u) \in \mathcal{K}_i$ implies that $(t, u) \in \mathcal{K}_i$.

$$(M, s) \models K_i\varphi \text{ if } (M, t) \models \varphi$$
$$\text{for all } t \text{ such that } (s, t) \in \mathcal{K}_i$$
$$(M, s) \models A_i\varphi \text{ if } \varphi \in \mathcal{A}_i(s)$$
$$(M, s) \models X_i\varphi \text{ if } (M, s) \models A_i\varphi \text{ and } (M, s) \models K_i\varphi.$$

A formula $\varphi$ is said to be *valid* in awareness structure $M$, written $M \models \varphi$, if $(M, s) \models \varphi$ for all $s \in S$. A formula is valid in a class $\mathcal{N}$ of awareness structures, written $\mathcal{N} \models \varphi$, if it is valid for all awareness structures in $\mathcal{N}$, that is, if $N \models \varphi$ for all $N \in \mathcal{N}$.

Consider the following set of well-known axioms and inference rules:

Prop. All substitution instances of valid formulas of propositional logic.

K. $(K_i\varphi \land K_i(\varphi \Rightarrow \psi)) \Rightarrow K_i\psi$.

T. $K_i\varphi \Rightarrow \varphi$.

4. $K_i\varphi \Rightarrow K_iK_i\varphi$.

5. $\neg K_i\varphi \Rightarrow K_i\neg K_i\varphi$.

A0. $X_i\varphi \Leftrightarrow K_i\varphi \land A_i\varphi$.

MP. From $\varphi$ and $\varphi \Rightarrow \psi$ infer $\psi$ (modus ponens).

Gen$_K$. From $\varphi$ infer $K_i\varphi$.

It is well known that the axioms T, 4, and 5 correspond to the requirements that the $\mathcal{K}_i$ relations are reflexive, transitive, and Euclidean, respectively. Let $\mathbf{K}_n$ be the axiom system consisting of the axioms Prop, K and rules MP, and Gen$_K$. The following result is well known (see, for example, [Fagin, Halpern, Moses, and Vardi 1995] for proofs).

**Theorem 2.1:** *Let $\mathcal{C}$ be a (possibly empty) subset of $\{T, 4, 5\}$ and let $C$ be the corresponding subset of $\{r, t, e\}$. Then $\mathbf{K}_n \cup \{A0\} \cup \mathcal{C}$ is a sound and complete axiomatization of the language $\mathcal{L}_n^{K,X,A}(\Phi)$ with respect to $\mathcal{M}_n^C(\Phi)$.*

## 3  A logic for reasoning about knowledge of unawareness

To allow reasoning about knowledge of unawareness, we extend the language $\mathcal{L}_n^{K,X,A}(\Phi)$ by adding a countable set of propositional variables $\mathcal{X} = \{x, y, z, \ldots\}$ and allowing universal quantification over these variables. Thus, if $\varphi$ is a formula, then so is $\forall x\varphi$. As usual, we take $\exists x\varphi$ to be an abbreviation for $\neg\forall x\neg\varphi$. Let $\mathcal{L}_n^{\forall,K,X,A}(\Phi, \mathcal{X})$ denote this extended language.

We assume that $\mathcal{X}$ is countably infinite for essentially the same reason that the set of variables is always taken to be infinite in first-order logic. Without it, we seriously limit the expressive power of the language. For example, a formula such as $\exists x\exists y(\neg(x \Leftrightarrow y) \land A_1x \land A_1y)$ says that there are two distinct formulas that agent 1 is aware of. We can similarly define formulas saying that there are $k$ distinct formulas that agent 1 is aware of. However, to do this we need $k$ distinct primitive propositions.

Essentially as in first-order logic, we can define inductively what it means for a variable $x$ to be *free* in a formula $\varphi$. If $\varphi$ does not contain the universal operator $\forall$, then every occurrence of $x$ in $\varphi$ is free; an occurrence of $x$ is free in $\neg\varphi$ (or $K_i\varphi$, $X_i\varphi$, $A_i\varphi$) iff the corresponding occurrence of $x$ is free in $\varphi$; an occurrence of $x$ in $\varphi_1 \wedge \varphi_2$ is free iff the corresponding occurrence of $x$ in $\varphi_1$ or $\varphi_2$ is free; and an occurrence of $x$ is free in $\forall y\varphi$ iff the corresponding occurrence of $x$ is free in $\varphi$ and $x$ is different from $y$. Intuitively, an occurrence of a variable is free in a formula if it is not bound by a quantifier. A formula that contains no free variables is called a *sentence*.

Let $\mathcal{S}_n^{\forall,K,X,A}(\Phi, \mathcal{X})$ denote the set of sentences in $\mathcal{L}_n^{\forall,K,X,A}(\Phi, \mathcal{X})$. If $\psi$ is a formula, let $\varphi[x/\psi]$ denote the formula that results by replacing all free occurrences of the variable $x$ in $\varphi$ by $\psi$. (If there is no free occurrence of $x$ in $\varphi$, then $\varphi[x/\psi] = \varphi$.) We extend this notion of substitution to sequences of variables, writing $\varphi[x_1/\psi_1, \ldots, x_n/\psi_n]$. We say that $\psi$ is *substitutable for $x$ in $\varphi$* if, for all propositional variables $y$, if an occurrence of $y$ is free in $\psi$, then the corresponding occurrence of $y$ is free in $\varphi[x/\psi]$.

We want to give semantics to formulas in $\mathcal{L}_n^{\forall,K,X,A}(\Phi, \mathcal{X})$ in awareness structures (where now we allow $\mathcal{A}_i(s)$ to be an arbitrary subset of $\mathcal{S}_n^{\forall,K,X,A}(\Phi, \mathcal{X})$). The standard approach for giving semantics to propositional quantification ([Engelhardt, Meyden, and Moses 1998; Kripke 1959; Bull 1969; Kaplan 1970; Fine 1970]) uses *semantic valuations*, much like in first-order logic. A *semantic valuation* $\mathcal{V}$ associates with each propositional variable and state a truth value, just as an interpretation $\pi$ associates with each primitive proposition and state a truth value. Then $(M, s, \mathcal{V}) \models x$ if $\mathcal{V}(x) = \textit{true}$ and $(M, s, \mathcal{V}) \models \forall x\varphi$ if $(M, s, \mathcal{V}') \models \varphi$ for all valuations $\mathcal{V}'$ that agree with $\mathcal{V}$ on all propositional variables other than $x$.[2] We write $\mathcal{V} \sim_x \mathcal{V}'$ if $\mathcal{V}(y) = \mathcal{V}'(y)$ for all variables $y \neq x$.

Using semantic valuations does not work in the presence of awareness. If $\mathcal{A}_i(s)$ consists only of sentences, then a formula such as $\forall x A_i x$ is guaranteed to be false since, no matter what the valuation is, $x \notin \mathcal{A}_i(s)$. The valuation plays no role in determining the truth of a formulas of the form $A_i\psi$. On the other hand, if we allow $\mathcal{A}_i(s)$ to include any formula in the language, then $(M, s, \mathcal{V}) \models \forall x A_i(x)$ iff $x \in \mathcal{A}_i(s)$. But then it is easy to check that $(M, s, \mathcal{V}) \models \exists x A_i(x)$ iff $x \in \mathcal{A}_i(s)$, which certainly does not seem to capture our intuition.

We want to interpret $\forall x A_i(x)$ as saying "for all sentences $\varphi \in \mathcal{S}_n^{\forall,K,X,A}(\Phi, \mathcal{X})$, $A_i(\varphi)$ holds". For technical reasons (which we explain shortly), we instead interpret this it as "for all formulas $\varphi \in \mathcal{L}_n^{K,X,A}(\Phi)$, $A_i(\varphi)$ holds". That is, we consider only sentences with no quantification. To achieve this, we use *syntactic valuations*, rather than *semantic valuations*. A *syntactic valuation* is a function $\mathcal{V} : \mathcal{X} \to \mathcal{L}_n^{K,X,A}(\Phi)$, which assigns to each variable a sentence in $\mathcal{L}_n^{K,X,A}(\Phi)$.

---

[2]We remark that the standard approach does not use separate propositional variables, but quantifies over primitive propositions. This makes it unnecessary to use valuations. It is easy to see that the definition we have given is equivalent to the standard definition. Using propositional variables is more convenient in our extension.

We give semantics to formulas in $\mathcal{L}_n^{\forall,K,X,A}(\Phi, \mathcal{X})$ by by induction on the total number of free and bound variables, with a subinduction on the length of the formula. The definitions for the constructs that already appear in $\mathcal{L}_n^{K,X,A}(\Phi)$ are the same. To deal with universal quantification, we just consider all possible replacements of the quantified variable by a sentence in $\mathcal{L}_n^{K,X,A}(\Phi)$.

- If $\varphi$ is a formula whose free variables are $x_1, \ldots, x_k$, then $(M, s, \mathcal{V}) \models \varphi$ if $(M, s, \mathcal{V}) \models \varphi[x_1/\mathcal{V}(x_1), \ldots, x_k/\mathcal{V}(x_k)]$

- $(M, s, \mathcal{V}) \models \forall x \varphi$ if $(M, s, \mathcal{V}') \models \varphi$ for all syntactic valuations $\mathcal{V}' \sim_x \mathcal{V}$.

Note that although $\varphi[x_1/\mathcal{V}(x_1), \ldots, \mathcal{V}(x_k)]$ may be a longer formula than $\varphi$, it involves fewer variables, since $\mathcal{V}(x_1), \ldots, \mathcal{V}(x_k)$ do not mention variables. This is why it is important that we quantify only over sentences in $\mathcal{L}_n^{K,X,A}(\Phi)$; if we were to quantify over all sentences in $\mathcal{S}_n^{\forall,K,X,A}(\Phi, \mathcal{X})$, then the semantics would not be well defined. For example, to determine the truth of $\forall xx$, we would have to determine the truth of $x[x/\forall xx] = \forall xx$. This circularity would make $\models$ undefined. In any case, given our restrictions, it is easy to show that $\models$ is well defined. Since the truth of a sentence is independent of a valuation, for a sentence $\varphi$, we write $(M, s) \models \varphi$ rather than $(M, s, V) \models \varphi$.

Under our semantics, the formula $K_i(\exists x(A_j x \wedge \neg A_i x))$ is consistent and that it can be true at state $s$ even though there might be no formula $\psi$ in $\mathcal{L}_n^{K,X,A}(\Phi)$ such that $K_i((A_j \psi \wedge \neg A_i \psi))$. This situation can happen if, at all states agent $i$ considers possible, agent $j$ is aware of something agent $i$ is not, but there is no one formula $\psi$ such that agent $j$ is aware of $\psi$ in all states agent $i$ considers possible and agent $i$ is not aware of $\psi$ in all such states. By way of contrast, if $\exists x K_i(A_j x \wedge \neg A_i x)$ is true at state $s$, then there is a formula $\psi$ such that $K_i(A_j \psi \wedge \neg A_i \psi)$ holds at $s$. The difference between $K_i \exists x(A_j x \wedge \neg A_i x)$ and $\exists x K_i(A_j x \wedge \neg A_i x)$ is essentially the same as the difference between $\exists x K_i \varphi$ and $K_i(\exists x \varphi)$ in first-order modal logic (see, for example, [Fagin, Halpern, Moses, and Vardi 1995] for a discussion).

The next example illustrates how the logic of knowledge of awareness can be used to capture some interesting situations.

**Example 3.1:** Consider an investor (agent 1) and an investment fund broker (agent 2). Suppose that we have two facts that are relevant for describing the situation: the NASDAQ index is more likely to increase than to decrease tomorrow ($p$), and Amazon will announce a huge increase in earnings tomorrow ($q$). Let $S = \{s\}$, $\pi(s,p) = \pi(s,q) = \textbf{true}$, $\mathcal{K}_i = \{(s,s)\}$, $\mathcal{A}_1(s) = \{p, \exists x(A_2 x \wedge \neg A_1 x)\}$, and $\mathcal{A}_2(s) = \{p, q, A_2 q, \neg A_1 q, A_2 q \wedge \neg A_1 q\}$. Thus, both agents explicitly know that the NASDAQ index is more likely to increase than to decrease tomorrow. However, the broker also explicitly knows that Amazon will announce a huge increase in earnings tomorrow. Furthermore, the broker explicitly knows that he (broker) is aware of this fact and the investor is not. On the other hand, the investor explicitly knows that there is something that the broker is aware of but he is not. That is,

$$(M, s, \mathcal{V}) \models X_1 p \wedge X_2 p \wedge X_2 q \wedge \neg X_1 q$$
$$\wedge X_2(A_2 q \wedge \neg A_1 q) \wedge X_1(\exists x(A_2 x \wedge \neg A_1 x)).$$

Since $X_2(A_2q \land \neg A_1q)$ implies $\exists x X_2(A_2x \land \neg A_1x)$, there is some formula $x$ such that the broker knows that the investor is unaware of $x$ although he is aware of $x$. However, since $(M, s, \mathcal{V}) \models \neg A_2(\exists x(A_2x \land \neg A_1x))$, it follows that $(M, s, \mathcal{V}) \models \neg X_2(\exists x(A_2x \land \neg A_1x))$. ∎

It may seem unreasonable that, in Example 3.1, the broker is aware of the formula $A_2q \land \neg A_1q$, without being aware of $\exists x(A_2x \land \neg A_1x)$. Of course, if the broker were aware of this formula, then $X_2((\exists x(A_2x \land \neg A_1x)))$ would hold at state $s$. This example suggests that we may want to require various properties of awareness. Here are some that are relevant in this context:

- Awareness is *closed under existential quantification* if $\varphi \in \mathcal{A}_i(s)$, $\varphi = \varphi'[x/\psi]$ and $\psi \in \mathcal{L}_n^{K,X,A}(\Phi)$, then $(\exists x \varphi') \in \mathcal{A}_i(s)$.

- Awareness is *generated by primitive propositions* if, for all agents $i$, $\varphi \in \mathcal{A}_i(s)$ iff all the primitive propositions that appear in $\varphi$ are in $\mathcal{A}_i(s) \cap \Phi$. That is, an agent is aware of $\varphi$ iff she is aware of all the primitive propositions that appear in $\varphi$.

- *Agents know what they are aware of* if, for all agents $i$ and all states $s, t$ such that $(s, t) \in \mathcal{K}_i$ we have that $\mathcal{A}_i(s) = \mathcal{A}_i(t)$.

Closure under existential quantification does *not* hold in Example 3.1. It is easy to see that it is a consequence of awareness being generated by primitive propositions. As shown by Halpern [2001] and Halpern and Rêgo [2005], a number of standard models of awareness in the economics literature (e.g., [Heifetz, Meier, and Schipper 2003; Modica and Rustichini 1999]) can be viewed as instances of the FH model where awareness is taken to be generated by primitive propositions and agents know what they are aware of. While assuming that awareness is generated by primitive propositions seems like quite a reasonable assumption if there is no existential quantification in the language, it does not seem quite so reasonable in the presence of quantification. For example, if awareness is generated by primitive propositions, then the formula $A_i(\exists x \neg A_i x)$ is valid, which does not seem to be reasonable in many applications. For some applications it may be more reasonable to instead assume only that awareness is *weakly generated by primitive propositions*. This is the case if, for all states $s$ and agents $i$,

- $\neg \varphi \in \mathcal{A}_i(s)$ iff $\varphi \in \mathcal{A}_i(s)$;

- $\varphi \land \psi \in \mathcal{A}_i(s)$ iff $\varphi, \psi \in \mathcal{A}_i(s)$;

- $K_i \varphi \in \mathcal{A}_i(s)$ iff $\varphi \in \mathcal{A}_i(s)$;

- $A_i \varphi \in \mathcal{A}_i(s)$ iff $\varphi \in \mathcal{A}_i(s)$;

- $X_i \varphi \in \mathcal{A}_i(s)$ iff $\varphi \in \mathcal{A}_i(s)$;

- if $\forall x \varphi \in \mathcal{A}_i(s)$, then $p \in \mathcal{A}_i(s)$ for each primitive proposition $p$ that appears in $\forall x \varphi$;

- if $\varphi[x/\psi] \in \mathcal{A}_i(s)$ for some formula $\psi \in \mathcal{L}_n^{K,X,A}(\Phi)$, then $\exists x \varphi \in \mathcal{A}_i(s)$.

If the language does not have quantification, then awareness is weakly generated by primitive propositions iff it is generated by primitive propositions. However, with quantification in the language, while it is still true that if awareness is generated by primitive propositions then it is weakly generated by primitive propositions, the converse does not necessarily hold. For example, if $\mathcal{A}_1(s) = \emptyset$ for all $s$, then awareness is weakly generated by primitive propositions. Intuitively, not being aware of any formulas is consistent with awareness being weakly generated by primitive propositions. However, if agent 1's awareness is generated by primitive propositions, then, for example, $\exists x A_j x$ must be in $\mathcal{A}_1(s)$ for all $s$ and all agents $j$.

## 4  Axiomatization

In this section, we provide a sound and complete axiomatization of the logic described in the previous section. We show that, despite the fact that we have a different language and used a different semantics for quantification, essentially the same axioms characterize our definition of quantification as those that have been shown to characterize the more traditional definition. Indeed, our axiomatization is very similar to the multi-agent version of an axiomatization given by Fine [1970] for a variant of his logic where the range of quantification is restricted.

Consider the following axioms for quantification:

$1_\forall$. $\forall x \varphi \Rightarrow \varphi[x/\psi]$ if $\psi$ is a quantifier-free formula substitutable for $x$ in $\varphi$.

$K_\forall$. $\forall x(\varphi \Rightarrow \psi) \Rightarrow (\forall x \varphi \Rightarrow \forall x \psi)$.

$N_\forall$. $\varphi \Rightarrow \forall x \varphi$ if $x$ is not free in $\varphi$.

Barcan. $\forall x K_i \varphi \Rightarrow K_i \forall x \varphi$.

$\text{Gen}_\forall$. From $\varphi$ infer $\forall x \varphi$.

These axioms are almost identical to the ones considered by Fine [1970], except that we restrict $1_\forall$ to quantifier-free formulas; Fine allows arbitrary formulas to be substituted (provided that they are substitutable for $x$). $K_\forall$ and $\text{Gen}_\forall$ are analogues to the axiom K and rule of inference $\text{Gen}_K$ in $\mathbf{K}_n$. The Barcan axiom, which is well-known in first-order modal logic, captures the relationship between quantification and $K_i$.

Let $\mathbf{K}_n^\forall$ be the axiom system consisting of the axioms in $\mathbf{K}_n$ together with $\{A0, 1_\forall, K_\forall, N_\forall,$ Barcan, $\text{Gen}_\forall\}$.

**Theorem 4.1:** *Let $\mathcal{C}$ be a (possibly empty) subset of $\{\text{T}, 4, 5\}$ and let $C$ be the corresponding subset of $\{r, t, e\}$. If $\Phi$ is countably infinite, then $\mathbf{K}_n^\forall \cup \mathcal{C}$ is a sound and complete axiomatization of the language $\mathcal{L}_n^{\forall, K, X, A}(\Phi, \mathcal{X})$ with respect to $\mathcal{M}_n^{\forall, C}(\Phi, \mathcal{X})$.*[3]

---

[3]We remark that Prior [1956] showed that, in the context of first-order modal logic, the Barcan axiom is not needed in the presence of the axioms of **S5** (that is $T$, 4, and 5). The same argument works here.

**Proof:** Showing that a provable formula $\varphi$ is valid can be done by a straightforward induction on the length of the proof of $\varphi$, using the fact that all axioms are valid in the appropriate set of models and all inference rules preserve validity.

In the standard completeness proof for modal logic, a canonical model $M^c$ is constructed where the states are maximal consistent sets of formulas. It is then shown that if $s_V$ is the state corresponding to the maximal consistent set $V$, then $(M^c, s_V) \models \varphi$ iff $\varphi \in V$. This will not quite work in our logic. We would need to define a canonical valuation function to give semantics for formulas containing free variables. We deal with this problem by considering states in the canonical model to consist of maximal consistent sets of sentences. There is another problem in the presence of quantification since there may be a maximal consistent set $V$ of sentences such that $\neg \forall x \varphi \in V$, but $\varphi[x/\psi]$ for all $\psi \in \mathcal{L}_n^{K,X,A}(\Phi)$. That is, there is no witness to the falsity of $\forall x \varphi$ in $V$. We deal with this problem by restricting to maximal consistent sets $V$ that are *acceptable* in the sense that if $\neg \forall x \varphi \in V$, then $\neg \varphi[x/q] \in V$ for some primitive proposition $q \in \Phi$. This argument requires $\Phi$ to be infinite. The details can be found in the appendix. ∎

To understand why, in general, we need to assume $\Phi$ is countably infinite, consider the case where there is only one agent and $\Phi = \{p\}$. Let $\varphi$ be the formula that essentially forces the S5 axioms to hold:
$$\forall x (Kx \Rightarrow KKx) \wedge (\neg Kx \Rightarrow K\neg Kx).$$
As is well-known, the S5 axioms force every formula in $\mathcal{L}_1^K$ to be equivalent to a depth-one formula (i.e., one without nested K's). Thus, it is not hard to show that there exists a finite set $F$ of formulas in $\mathcal{L}_1^{K,X,A}(\{p\})$ such that for all formulas $\psi$ with one free variable $y$ and no quantification, we have for $C \subseteq \{r, e, t\}$
$$\mathcal{M}_1^{\forall,C}(\Phi, \mathcal{X}) \models (\varphi \wedge \forall x Ax) \Rightarrow (\forall y \psi \Leftrightarrow \wedge_{\sigma \in F} \psi[y/\psi]).$$

Thus, if $\Phi$ has only one primitive proposition, then there are circumstances under which universal quantification is equivalent to a finite conjunction. We can construct similar examples if $\Phi$ is an arbitrary finite set of propositions even if there is more than one agent. (In the latter case, we add a formula $\varphi'$ to the antecedent that says that agent 1 knows that all agents have the same knowledge that he does: $\forall x K_1 \wedge_{i=2}^n (K_i x \Leftrightarrow K_1 x)$.) Thus, if $\Phi$ is finite, we would need extra axioms to capture the fact universal quantification can sometimes be equivalent to a finite conjunction. We remark that this phenomenon of needing additional axioms if $\Phi$ is finite has been observed before in the literature (cf. [Fagin, Halpern, and Vardi 1992; Halpern and Lakemeyer 2001]).

If we make further assumptions about the awareness operator, these can also be captured axiomatically. For example, as shown by FH, the assumption that agents know what they are aware of corresponds to the axioms
$$A_i \varphi \Rightarrow K_i A_i \varphi \text{ and}$$
$$\neg A_i \varphi \Rightarrow K_i \neg A_i \varphi.$$

It is not hard to check that awareness being generated by primitive propositions can be captured by the following axiom:

$$A_i\varphi \Leftrightarrow \wedge_{\{p\in\Phi:\ p\text{ occurs in }\varphi\}} A_i p.$$

In this axiom, the empty conjunction is taken to be vacuously true, so that $A_i\varphi$ is vacuously true if no primitive propositions occur in $\varphi$.

We can axiomatize the fact that awareness is weakly generated by primitive propositions using the following axioms:

A1. $A_i(\varphi \wedge \psi) \Leftrightarrow A_i\varphi \wedge A_i\psi$.

A2. $A_i\neg\varphi \Leftrightarrow A_i\varphi$.

A3. $A_i X_j \varphi \Leftrightarrow A_i\varphi$.

A4. $A_i A_j \varphi \Leftrightarrow A_i\varphi$.

A5. $A_i K_j \varphi \Leftrightarrow A_i\varphi$.

A6. $A_i\varphi \Rightarrow A_i p$ if $p \in \Phi$ occurs in $\varphi$.

A7. $A_i\varphi[x/\psi] \Rightarrow A_i\exists x\varphi$, where $\psi \in \mathcal{L}_n^{K,X,A}(\Phi)$.

As noted in [Fagin, Halpern, Moses, and Vardi 1995], the first five axioms capture awareness generated by primitive propositions in the language $\mathcal{L}_n^{K,X,A}(\Phi)$; we need A6 and A7 to deal with quantification. A7 captures the fact that awareness is closed under existential quantification.

## 5 Complexity

Since the logic is axiomatizable, the validity problem is at worst recursively enumerable. As the next theorem shows, the validity problem is no better than r.e.

**Theorem 5.1:** *The problem of deciding if a formula in the language $\mathcal{L}_n^{\forall,K,X,A}(\Phi, \mathcal{X})$ is valid in $\mathcal{M}_n^C(\Phi, \mathcal{X})$ is r.e.-complete, for all $C \subseteq \{r, t, e\}$ and $n \geq 1$.*

**Proof:** The fact that deciding validity is r.e. follows immediately from Theorem 2.1. For the hardness result, we show that, for every formula $\varphi$ in first-order logic over a language with a single binary predicate can be translated to a formula $\varphi^t \in \mathcal{L}_1^{\forall,K,A}(\Phi, \mathcal{X})$ such that $\varphi$ is valid over relational models iff $\varphi^t$ is valid in $\mathcal{M}_n^\emptyset(\Phi, \mathcal{X})$ (and hence in $\mathcal{M}_n^C(\Phi, \mathcal{X})$, for all $C \subseteq \{r, t, e\}$. We leave details of the reduction to the appendix. The result follows from the well-known fact that the validity problem for first-order logic with one binary predicate is r.e. ∎

Theorem 5.1 is somewhat surprising, since Fine [1970] shows that his logic (which is based on S5) is decidable. It turns out that each of the following suffices to get undecidability: (a) the presence of the awareness operator, (b) the presence of more than one agent, or (c) not having $e \in C$ (i.e., not assuming that the $\mathcal{K}$ relation satisfies the Euclidean property). The fact that awareness gives undecidability is the content of Theorem 5.1; Theorem 5.2 shows that having $n \geq 2$ or $e \notin C$ suffices for undecidability as well. On the other hand, Theorem 5.3 shows that if $n = 1$ and $e \in C$, then the problem is decidable. Although, as we have observed, our semantics is slightly differently from that of Fine, we believe that corresponding results hold in his setting. Thus, he gets decidability because he does not have awareness, restricts to a single agent, and considers S5 (as opposed to say, S4).

Let $\mathcal{L}_n^{\forall,K}(\Phi, \mathcal{X})$ consist of all formulas in $\mathcal{L}_n^{\forall,K,X,A}(\Phi, \mathcal{X})$ that do not mention the $A_i$ or $X_i$ operators.

**Theorem 5.2**: *The problem of deciding if a formula in the language $\mathcal{L}_n^{\forall,K}(\Phi)$ is valid in $\mathcal{M}_n^C(\Phi)$ is r.e.-complete if $n \geq 2$ or if $e \notin C$.*

**Theorem 5.3**: *The validity problem for the language $\mathcal{L}_1^{\forall,K}(\Phi, \mathcal{X})$ with respect to the structures in $\mathcal{M}_1^C(\Phi, \mathcal{X})$ for $C \supseteq \{e\}$ is decidable.*

Interestingly, the role of the Euclidean property in these complexity results mirrors its role in complexity for $\mathcal{L}_n^K$, basic epistemic logic without awareness or quantification. As we have shown [Halpern and Rêgo 2006a], the problem of deciding if a formula in the language $\mathcal{L}_n^K(\Phi)$ is valid in $\mathcal{M}_n^C(\Phi)$ is PSPACE complete if $n \geq 2$ or $n \geq 1$ and $e \notin C$; if $n = 1$ and $e \in C$, it is co-NP-complete.

## 6 Conclusion

We have proposed a logic to model agents who are able to reason about their lack of awareness. We have shown that such reasoning arises in a number of situations. We have provided a complete axiomatization for the logic, and examined the complexity of the validity problem.

Our original motivation for considering knowledge of unawareness came from game theory. Notions like Nash equilibrium do not make sense in the presence of lack of awareness. Intuitively, a set of strategies is a Nash equilibrium if each agent would continue playing the same strategy despite knowing what strategies the other agents are using. But if an agent is not aware of the moves available to other agents, then he cannot even contemplate the actions of other players. In a companion paper [Halpern and Rêgo 2006b], we show how to generalize the notion of Nash equilibrium so that it applies in the presence of (knowledge of) unawareness. Feinberg [2004] has already shown that awareness can play a significant role in analyzing games. (In particular, he shows that a small probability of an agent not being aware of the possibility of defecting in finitely repeated prisoners dilemma can lead to cooperation.) It is not hard to show that knowledge of unawareness can have a similarly significant impact. Consider, for example, a chess game. If we interpret "lack of awareness" as "unable to compute" (cf. [Fagin and Halpern 1988]), then although all players understand in principle all the moves that can be made, they are certainly not aware of all consequences of all moves. Such lack of awareness has strategic implications. For example, in cases where the opponent is under time pressure, experts will make deliberately make moves that lead to positions that are hard to analyze. (In our language, these are positions where there is a great deal of unawareness.) The logic we have presented here provides an initial step to modeling the reasoning that goes on in games with (knowledge of) unawareness. To fully model what is going on, we need to capture probability as well as awareness and knowledge. We do not think that there will be any intrinsic difficulty in extending our logic to handle probability along the lines of the work in [Fagin and Halpern 1994; Fagin, Halpern, and Megiddo 1990], although we have not checked the details.

## A  Proof of Theorems

**Theorem 4.1:** *Let $\mathcal{C}$ be a (possibly empty) subset of $\{\mathrm{T}, 4, 5\}$ and let $C$ be the corresponding subset of $\{r, t, e\}$. Then $\mathrm{K}_n^\forall \cup \mathcal{C}$ is a sound and complete axiomatization of the language $\mathcal{L}_n^{\forall, K, X, A}(\Phi, \mathcal{X})$ with respect to $\mathcal{M}_n^{\forall, C}(\Phi, \mathcal{X})$.*

**Proof:** We give the proof only for the case $\mathcal{C} = \emptyset$; the other cases follow using standard techniques (see, for example, [Fagin, Halpern, Moses, and Vardi 1995; Hughes and Cresswell 1996]). Showing that a provable formula $\varphi$ is valid can be done by a straightforward induction on the length of the proof of $\varphi$, using the fact that all axioms in $\mathrm{K}_n^\forall$ are valid in $\mathcal{M}_n^{\forall, \emptyset}(\Phi, \mathcal{X})$ and all inference rules preserve validity in $\mathcal{M}_n^{\forall, \emptyset}(\Phi, \mathcal{X})$.

As we said in the main text, we prove completeness by modifying the standard canonical model construction, restricting to acceptable maximal consistent sets of sentences. Thus, the first step in the proof is to guarantee that every consistent sentence is included in an acceptable maximal consistent set of sentences.

If $q$ is a primitive proposition, we define $\varphi[q/x]$ and the notion of $x$ being substitutable for $q$ just as we did for the case that $q$ is a propositional variable.

**Lemma A.1:** *If $\mathrm{K}_n^\forall \cup \mathcal{C} \vdash \varphi$ and $x$ is substitutable for $q$ in $\varphi$, then $\mathrm{K}_n^\forall \cup \mathcal{C} \vdash \forall x \varphi[q/x]$.*

**Proof:** We first show by induction on the length of the proof of $\varphi$ that if $z$ is a variable that does not appear in any formula in the proof of $\varphi$, then $\mathrm{K}_n^\forall \cup \mathcal{C} \vdash \varphi[q/z]$. If there is a proof of $\varphi$ of length one, then $\varphi$ is an instance of an axiom. It is easy to see that $\varphi[q/z]$ is an instance of the same axiom. (We remark that it is important in the case of axioms $\mathrm{N}_\forall$ and $1_\forall$ that $z$ does not occur in $\varphi$.) Suppose that the lemma holds for all $\varphi'$ that have a proof of length no greater than $k$, and suppose that $\varphi$ has a proof of length $k+1$ where $z$ does not occur in any formula of the proof. If the last step of the proof of $\varphi$ is an axiom, then $\varphi$ is an instance of an axiom, and we have already dealt with this case. Otherwise, the last step in the proof of $\varphi$ is an application of either MP, $\mathrm{Gen}_K$, or $\mathrm{Gen}_\forall$. We consider these in turn.

If MP is applied at the last step, then there exists some $\varphi'$, such that $\varphi'$ and $\varphi' \Rightarrow \varphi$ were previously proved and, by assumption, $z$ does not occur in any formula of their proof. By the induction hypothesis, both $\varphi'[q/z]$ and $(\varphi' \Rightarrow \varphi)[q/z] = \varphi'[q/z] \Rightarrow \varphi[q/z]$ are provable. The result now follows by an application of MP.

The argument for $\text{Gen}_K$ and $\text{Gen}_\forall$ is essentially identical, so we consider them together. Suppose that $\text{Gen}_K$ (resp., $\text{Gen}_\forall$) is applied at the last step. Then $\varphi$ has the form $K_i\varphi'$ (resp., $\forall y \varphi'$) and there is a proof of length at most $k$ for $\varphi'$ where $z$ does not occur in any formula in the proof. Thus, by the induction hypothesis, $\varphi'[q/z]$ is provable. By applying $\text{Gen}_K$ (resp., $\text{Gen}_\forall$), it immediately follows that $\varphi[q/z]$ is provable.

This completes the proof that $\varphi[q/z]$ is provable. By applying $\text{Gen}_\forall$, it follows that $\forall z \varphi[q/z]$ is provable. Since $x$ is substitutable for $q$ in $\varphi$, $x$ must be substitutable for $z$ in $\varphi[q/z]$. Thus, by applying the axiom $1_\forall$ and MP, we can prove $\varphi[q/x]$. The fact that $\forall x \varphi[q/x]$ is provable now follows from $\text{Gen}_\forall$. ∎

**Lemma A.2:** *Every $\mathbf{K}_n^\forall$-consistent sentence $\varphi$ is contained in some acceptable maximal $\mathbf{K}_n^\forall$-consistent set of sentences.*

**Proof:** We first show that if $\Delta$ is a finite $\mathbf{K}_n^\forall$-consistent set of sentences, $\neg \forall x \psi \in \Delta$, and $q$ is a primitive proposition that does not appear in any sentence in $\Delta$, then $\Delta \cup \neg \varphi[x/q]$ is $\mathbf{K}_n^\forall$-consistent. Suppose not. Then there exist sentences $\beta_1, \ldots, \beta_k \in \Delta$ such that

$$\mathbf{K}_n^\forall \vdash (\beta_1 \wedge \ldots \wedge \beta_k) \Rightarrow \psi[x/q].$$

By Lemma A.1, we have

$$\mathbf{K}_n^\forall \vdash \forall x((\beta_1 \wedge \ldots \wedge \beta_k) \Rightarrow \psi). \tag{1}$$

Now applying $K_\forall$ and $N_\forall$, and using the fact that $x$ is not free in $\beta_1 \wedge \ldots \wedge \beta_k$ (since $\beta_1, \ldots, \beta_k$ are sentences), it easily follows that

$$\mathbf{K}_n^\forall \vdash (\beta_1 \wedge \ldots \wedge \beta_k) \Rightarrow \forall x \psi.$$

Since $\beta_1, \ldots, \beta_k, \neg \forall x \psi \in \Delta$, this contradicts the assumption that $\Delta$ is $\mathbf{K}_n^\forall$-consistent.

We now use standard techniques to construct an acceptable maximal $\mathbf{K}_n^\forall$-consistent set of sentences containing $\varphi$. Consider an enumeration $\{\psi_1, \psi_2, \ldots\}$ of the sentences in $\mathcal{S}_n^{\forall,K,X,A}$. We construct a sequence of $\mathbf{K}_n^\forall$-consistent sets of sentences $\Delta_0, \Delta_1, \ldots$. Let $\Delta_0 = \{\varphi\}$. For all $k \geq 1$, if $\Delta_{k-1} \cup \{\psi_k\}$ is not $\mathbf{K}_n^\forall$-consistent, $\Delta_k = \Delta_{k-1}$. If $\Delta_{k-1} \cup \{\psi_k\}$ is $\mathbf{K}_n^\forall$-consistent, if $\psi_k$ is not of the form $\neg \forall y \psi'$, then $\Delta_k = \Delta_{k-1} \cup \{\psi_k\}$, while if $\psi_k$ is of the form $\neg \forall y \psi'$, then $\Delta_k = \Delta_{k-1} \cup \{\psi_k, \neg \psi'[y/q]\}$ for some $q \in \Phi$ not occurring in $\Delta_{k-1} \cup \{\psi_k\}$. By our earlier argument, it easily follows that each set $\Delta_k$ is $\mathbf{K}_n^\forall$-consistent. Let $\Delta$ be the union of the $\Delta_n$'s. It is easy to see that $\Delta$ is an acceptable maximal $\mathbf{K}_n^\forall$-consistent set of sentences that contains $\varphi$. ∎

For a set $\Gamma$ of formulas, define $\Gamma/K_i = \{\psi : K_i \psi \in \Gamma\}$.

**Lemma A.3:** *If $\Gamma$ is a $\mathbf{K}_n^\forall$-consistent set of sentences containing $\neg K_i\varphi$, then $\Gamma/K_i \cup \{\neg\varphi\}$ is $\mathbf{K}_n^\forall$-consistent.*

**Proof:** This is a standard modal logic argument; see, for example, [Hughes and Cresswell 1996, Lemma 6.4]. We omit details here. ∎

**Lemma A.4:** *If $\Gamma$ is an acceptable maximal $\mathbf{K}_n^\forall$-consistent set of sentences and $\neg K_i\varphi \in \Gamma$, then there exists an acceptable maximal $\mathbf{K}_n^\forall$-consistent set of sentences $\Delta$ such that $(\Gamma/K_i \cup \{\neg\varphi\}) \subseteq \Delta$.*

**Proof:** We first show that if $\Gamma/K \cup \{\gamma_1, \ldots, \gamma_n, \neg\forall x\psi\}$ is a $\mathbf{K}_n^\forall$-consistent set of sentences, then there exists some $q \in \Phi$ such that $\Gamma/K_i \cup \{\gamma_1, \ldots, \gamma_n, \neg\forall x\psi, \neg\psi[x/q]\}$ is $\mathbf{K}_n^\forall$-consistent. For suppose not. Then, for all $q$, there must exist $\beta_1, \ldots, \beta_k \in \Gamma/K_i$ such that

$$\mathbf{K}_n^\forall \vdash (\beta_1 \wedge \ldots \wedge \beta_k) \Rightarrow (\gamma \Rightarrow \psi[x/q]),$$

where $\gamma = \gamma_1 \wedge \ldots \wedge \gamma_n \wedge \neg\forall x\psi$. By $\text{Gen}_K$, we have that

$$\mathbf{K}_n^\forall \vdash K_i((\beta_1 \wedge \ldots \wedge \beta_k) \Rightarrow (\gamma \Rightarrow \psi[x/q])).$$

Applying axiom K, and using the fact that $K(\beta_1 \wedge \ldots \wedge \beta_k) \in \Gamma$ and $\Gamma$ is maximal, we have that $K_i(\gamma \Rightarrow \psi[x/q]) \in \Gamma$ for all $q$. Since $\Gamma$ is acceptable, it must be the case that $\forall x K_i(\gamma \Rightarrow \psi) \in \Gamma$. By the Barcan axiom, it follows that $K_i \forall x(\gamma \Rightarrow \psi) \in \Gamma$. Since $\gamma$ is a sentence, applying $K_\forall$, $N_\forall$, K, and $\text{Gen}_K$, it follows that $K_i \forall x(\gamma \Rightarrow \psi) \Rightarrow K_i(\gamma \Rightarrow \forall x\psi)$ is provable in $\mathbf{K}_n^\forall$. Hence, $K_i(\gamma \Rightarrow \forall x\psi) \in \Gamma$. Thus, $\gamma \Rightarrow \forall x\psi \in \Gamma/K_i$. But this contradicts the consistency of $\Gamma/K \cup \{\gamma_1, \ldots, \gamma_n, \neg\forall x\psi\}$.

We now proceed much as in the proof of Lemma A.2. Given an enumeration $\psi_1, \psi_2, \ldots$ of the sentences in $\mathcal{S}_n^{\forall,K,X,A}$, we construct a sequence of $\mathbf{K}_n^\forall$-consistent sets of sentences $\Delta_0, \Delta_1, \ldots$. Let $\Delta_0 = \Gamma/K_i \cup \{\neg\varphi\}$. (Note that Lemma A.3 implies that $\Delta_0$ is $\mathbf{K}_n^\forall$-consistent.) For all $k \geq 1$, if $\Delta_{k-1} \cup \{\psi_k\}$ is not $\mathbf{K}_n^\forall$-consistent, $\Delta_k = \Delta_{k-1}$. If $\Delta_{k-1} \cup \{\psi_k\}$ is $\mathbf{K}_n^\forall$-consistent, if $\psi_k$ is not of the form $\neg\forall y\psi'$, then $\Delta_k = \Delta_{k-1} \cup \{\psi_k\}$, while if $\psi_k$ is of the form $\neg\forall y\psi'$, then $\Delta_k = \Delta_{k-1} \cup \{\psi_k, \neg\psi'[y/q]\}$ for some $q \in \Phi$ such that $\Delta_{k-1} \cup \{\psi_k, \neg\psi'[y/q]\}$ is consistent. (Such a $q$ exists by our earlier argument.) It is easy to see that $\Delta = \cup_n \Delta_n$ is the desired acceptable maximal $\mathbf{K}_n^\forall$-consistent set of sentences. ∎

We are now able to prove the following key lemma.

**Lemma A.5:** *If $\varphi$ is a $\mathbf{K}_n^\forall$-consistent sentence, then $\varphi$ is satisfiable in $\mathcal{M}_n^{\forall,\emptyset}(\Phi, \mathcal{X})$.*

**Proof:** Let $M^c = (S, \mathcal{K}_1, \ldots, \mathcal{K}_n, \mathcal{A}_1, \ldots, \mathcal{A}_n, \pi)$ be a canonical awareness structure constructed as follows

- $S = \{s_V : V \text{ is an acceptable maximal } \mathbf{K}_n^\forall\text{-consistent set of sentences}\};$

- $\pi(s_V, p) = \begin{cases} 1 & \text{if } p \in V, \\ 0 & \text{if } p \notin V; \end{cases}$

- $\mathcal{A}_i(s_V) = \{\varphi : A_i\varphi \in V\};$

- $\mathcal{K}_i(s_V) = \{s_W : V/K_i \subseteq W\}.$

We show as usual that if $\psi$ is a sentence, then

$$(M^c, s_V) \models \psi \text{ iff } \psi \in V. \tag{2}$$

Note that this claim suffices to prove Lemma A.5 since, if $\varphi$ is a $\mathbf{K}_n^\forall$-consistent sentence, by Lemma A.2, it is contained in an acceptable maximal $\mathbf{K}_n^\forall$-consistent set of sentences.

We prove (2) by induction of the depth of nesting of $\forall$, with a subinduction on the length of the sentence.

The base case is if $\psi$ is a primitive proposition, in which case (2) follows immediately from the definition of $\pi$. For the inductive step, given $\psi$, suppose that (2) holds for all formulas $\psi'$ such that either the depth of nesting for $\forall$ in $\psi'$ is less than that in $\psi$, or the depth of nesting is the same, and $\psi'$ is shorter than $\psi$. We proceed by cases on the form of $\psi$.

- If $\psi$ has the form $\neg\psi'$ or $\psi_1 \wedge \psi_2$, then the result follows easily from the inductive hypothesis.

- If $\psi$ has the form $A_i\psi'$, then note that $\varphi'$ is a sentence and $(M^c, s_V) \models A_i\psi'$ iff $\psi' \in \mathcal{A}_i(s_V)$ iff $A_i\psi' \in V$.

- If $\psi$ has the form $K_i\psi'$, then if $\psi \in V$, then $\psi' \in W$ for every $W$ such that $s_W \in \mathcal{K}_i(s_V)$. By the induction hypothesis, $(M^c, s_W) \models \psi'$ for every $s_W \in \mathcal{K}_i(s_V)$, so $(M^c, s_V) \models K_i\psi'$. If $\psi \notin V$, then $\neg\psi \in V$ since $V$ is a maximal $\mathbf{K}_n^\forall$-consistent set. By Lemma A.4, there exists an acceptable maximal $\mathbf{K}_n^\forall$-consistent set of sentences $W$ such that $(V/K_i \cup \{\neg\psi'\}) \subseteq W$. By the induction hypothesis, $(M^c, s_W) \not\models \psi'$. Thus, $(M^c, s_V) \not\models K_i\psi'$.

- If $\psi$ has the form $X_i\psi'$, the argument is immediate from the preceding two cases and the observation that $(M, s_V) \models X_i\psi'$ iff both $(M, s_V) \models K_i\psi'$ and $(M, s_V) \models A'_\psi$, while $X_i\psi' \in V$ iff both $K_i\psi' \in V$ and $A_i\psi' \in V$.

- Finally, suppose that $\psi = \forall x \psi'$. If $\psi \in V$ then, by axiom $1_\forall$, $\psi'[x/\varphi] \in V$ for all $\varphi \in \mathcal{L}_n^{K,X,A}(\Phi)$. The depth of nesting of $\psi'[x/\varphi]$ is less than that of $\forall x \psi'$, so by the induction hypothesis $(M, s_V) \models \psi'[x/\varphi]$ for all $\varphi \in \mathcal{L}_n^{K,X,A}(\Phi)$. By definition, $(M, s_V) \models \psi$, as desired. If $\psi \notin V$ then, since $V$ is acceptable, there exists a primitive proposition $q \in \Phi$ such that $\psi'[x/q] \notin V$. By the induction hypothesis, $(M^c, s_V) \not\models \psi'[x/q]$. Thus, $(M^c, s_V) \not\models \psi$, as desired.

To finish the completeness proof, suppose that $\varphi$ is valid in $\mathcal{M}_n^{\forall,\emptyset}(\Phi, \mathcal{X})$. Then, consider two cases: (1) $\varphi$ is a sentence, and (2) $\varphi$ is not a sentence. If (1), then $\neg\varphi$ is a sentence and is not satisfiable in $\mathcal{M}_n^{\forall,\emptyset}(\Phi, \mathcal{X})$. So, by Lemma A.5, $\neg\varphi$ is not $\mathbf{K}_n^\forall$-consistent. Thus, $\varphi$ is provable in $\mathbf{K}_n^\forall$. If (2) and $\{x_1, \ldots, x_k\}$ is the set of free variables in $\varphi$, then $\forall x_1 \ldots \forall x_k \varphi$ is a valid sentence. Thus, by case (1), $\forall x_1 \ldots \forall x_k \varphi$ is provable in $\mathbf{K}_n^\forall$. Applying $1_\forall$ repeatedly it follows that $\varphi$ is provable in $\mathbf{K}_n^\forall$, as desired. ∎

**Theorem 5.1:** *Deciding if a formula in the language $\mathcal{L}_n^{\forall,K,X,A}(\Phi, \mathcal{X})$ is valid in $\mathcal{M}_n^C(\Phi, \mathcal{X})$ is r.e.-complete, for all $C \subseteq \{r, t, e\}$ and $n \geq 1$.*

**Proof:** The fact that deciding validity is r.e. follows immediately from Theorem 2.1. For the hardness result, to do the reduction, we first fix some notation. Take an *R-formula* to be a first-order formula (without equality) whose only nonlogical symbol is the binary predicate $R$. Take an *R-model* to be a relational structure which provides an interpretation for $R$. A *countable R model* is an $R$ model with a countable domain. It is well known that the satisfiability problem for $R$-formulas is undecidable [Lewis 1979]. Thus, it suffices to reduce the satisfiability problem for $R$-formulas to the satisfiability problem for formulas in $\mathcal{L}_n^{\forall,K,X,A}(\Phi, \mathcal{X})$.

For easy of exposition, assume that the set $\Phi$ of primitive propositions includes $q_1$, $q_2$, and $r$; later we show how to get rid of this assumption. Given an $R$-model $N$, we will construct an awareness structure $M$ that represents $N$. Roughly speaking, a state in $M$ represents an ordered pair of domain elements in $N$. The primitive proposition $r$ will be true at a state $s$ in $M$ iff $R(d_1, d_2)$ is true in $N$ of the pair $(d_1, d_2)$ represented by $s$. The primitive propositions $q_1$ and $q_2$ are used to encode $d_1$ and $d_2$. Let $\sigma$ be the awareness formula that, roughly speaking, forces it to be the case that for all states $s$, if $r$ is true at some state $t$ that represents $(d_1, d_2)$ such that $(s, t) \in \mathcal{K}$, then $r$ is true at all states $t'$ that represent $(d_1, d_2)$ such that $(s, t') \in \mathcal{K}$. (It follows that if $\neg r$ is true at some state $t$ that represents $(d_1, d_2)$ such that $(s, t) \in \mathcal{K}$, then $\neg r$ is true at all states $t'$ that represent $(d_1, d_2)$ such that $(s, t') \in \mathcal{K}$. The formula $\sigma$ is

$$\forall x_1 \forall x_2 (\neg K \neg (A(x_1 \wedge q_1) \wedge A(x_2 \wedge q_2) \wedge r) \Rightarrow K((A(x_1 \wedge q_1) \wedge A(x_2 \wedge q_2)) \Rightarrow r)).$$

Now we translate an $R$-formula $\psi$ to an awareness formula $\psi^t$. We consider only $R$-formulas formulas in *negation normal form*, i.e., formulas $\psi$ that use $\wedge, \vee, \forall$, and $\exists$, where the negation has been pushed in so that it occurs only in front of the predicate $R$. It is well known that every $R$-formula is equivalent to a formula in negation normal form.

- $(R(x, y))^t = \neg K \neg (r \wedge A(x \wedge q_1) \wedge A(y \wedge q_2))$
- $(\neg R(x, y))^t = \neg K \neg (\neg r \wedge A(x \wedge q_1) \wedge A(y \wedge q_2))$
- $(\varphi_1 \wedge \varphi_2)^t = \varphi_1^t \wedge \varphi_2^t$
- $(\varphi_1 \vee \varphi_2)^t = \varphi_1^t \vee \varphi_2^t$

- $(\forall x \varphi)^t = \forall x \varphi^t$

- $(\exists x \varphi)^t = \exists x \varphi^t$

We say that an awareness structure $M = (S, \mathcal{K}, \mathcal{A}, \pi)$ is *universal* if $\mathcal{K} = S \times S$. It is easy to see that if $M$ is a universal structure, then $M \in \mathcal{M}_1^C(\Phi, \mathcal{X})$ for all $C \subseteq \{r, t, e\}$. Moreover, an easy argument by induction on structure, whose proof we leave to the reader, shows the following.

**Lemma A.6:** *If $M = (S, \ldots)$ is a universal structure, then for all $R$-formulas in negation normal form and $\mathcal{V}$ is a syntactic valuation, then $(M, s, \mathcal{V}) \models \psi^t$ for some $s \in S$ iff $(M, s', \mathcal{V}) \models \psi^t$ for all states $s' \in S$.*

We write $(M, \mathcal{V}) \models \psi^t$ if $(M, s, \mathcal{V}) \models \psi^t$ for all $s \in S$.

Theorem 5.1 follows from the following claim:

For all $C \subseteq \{r, t, e\}$, $\varphi$ is satisfiable in an $R$-model iff $\varphi^t \wedge \sigma$ is satisfiable in $\mathcal{M}_1^C(\Phi, \mathcal{X})$. (3)

To prove (3), first suppose that $\psi$ is a satisfiable $R$-formula. It is well known that an $R$-formula is satisfiable iff it is satisfiable in a countable $R$-model [Enderton 1972] (that is, an $R$-model with a countable domain. (Of course, this result holds for arbitrary first-order formulas, not just $R$-formulas.) Thus, we can assume without loss of generality that $\psi$ is satisfied in the $R$-model $N$ with countable domain $D_N$.

Let $L$ be a surjection from $\mathcal{L}_1^{K,X,A}(\Phi)$ to $D_N$. (Since $D_N$ is countable, such a surjection exists.) Given the $R$-model $N$ with countable domain $D_N$, define $M_N = (S, \mathcal{K}, \mathcal{A}, \pi)$ to be the universal awareness structure such that

- $S = \{(d_1, d_2) : d_1, d_2 \in D_N\}$;

- $\pi((d_1, d_2), r) = \textbf{true}$ iff $(d_1, d_2) \in R$;

- $\pi((d_1, d_2), q) = \textbf{true}$ for all $q \in \Phi - \{r\}$;

- $\mathcal{A}((d_1, d_2)) = \{\psi \wedge q_1 : L(\psi) = d_1\} \cup \{\psi' \wedge q_2 : L(\psi') = d_2\}$.

It is easy to check that $M_N \models \sigma$; we leave the proof to the reader. Thus, it suffices to show that there is some state $s$ and syntactic valuation $\mathcal{V}$ such that $(M_N, s, \mathcal{V}) \models \varphi^t$. This follows from the following result.

**Lemma A.7:** *For every first-order formula $\psi$ in negation normal form, if $N \models \psi$ then $M_N \models \psi^t$.*

**Proof:** We actually prove a slightly more general result. A syntactic valuation $\mathcal{V}$ is *L-compatible* with a valuation $V$ on $N$ (that is, a function mapping variables to elements of $D_N$) if, for all variables $x$, $L(\mathcal{V}(x)) = V(x)$. We show that for all first-order formulas $\psi$ (not necessarily a sentence) and all valuations $V$ on $N$, if $(N,V) \models \psi$, then $(M_N, \mathcal{V}) \models \psi^t$ for all syntactic valuations $\mathcal{V}$ $L$-compatible with $V$. The proof is by induction on structure.

Suppose that $\psi = R(x,y)$. Then, $(N,V) \models \psi$ iff $s = (V(x), V(y)) \in R$. By definition, $s \in R$ iff $\pi(s,r) = \textbf{true}$ and $(M_N, s, \mathcal{V}) \models A(x \wedge q_1) \wedge A(y \wedge q_2)$ for all syntactic valuations $\mathcal{V}$ $L$-compatible with V. Since $M_N$ is universal and $R(x,y)^t = \neg K \neg (r \wedge A(x \wedge q_1) \wedge A(y \wedge q_2))$, it follows that $(M_N, \mathcal{V}) \models R(x,y)^t$ for all $\mathcal{V}$ $L$-compatible with $V$. A similar argument applies if $\psi$ is of the form $\neg R(x,y)$. If $\psi = \psi_1 \wedge \psi_2$ or $\psi = \psi_1 \vee \psi_2$, the result follows easily from the induction hypothesis. Suppose that $\psi = \forall x \psi'$, $(N,V) \models \forall x \psi'$, and $\mathcal{V}$ is $L$-compatible with $V$. We want to show that $(M_N, \mathcal{V}) \models \psi^t$. Since $\psi^t = \forall x(\psi')^t$, we must show that $(M_N, \mathcal{V}') \models (\psi')^t$ for all $\mathcal{V}' \sim_x \mathcal{V}$. Given a valuation $\mathcal{V}' \sim_x \mathcal{V}$, consider the valuation $V' = L \circ \mathcal{V}'$ on $N$; that is, $V'(y) = L(\mathcal{V}'(y))$ for all variables $y$. Clearly, if $y \neq x$, $V'(y) = L(\mathcal{V}'(y)) = L(\mathcal{V}(y)) = V(y)$. Thus, $V' \sim_x V$, so $(N, V') \models \psi'$. Moreover, since $\mathcal{V}'$ is clearly $L$-compatible with $V'$, it follows from the induction hypothesis that $(M_N, \mathcal{V}') \models (\psi')^t$. Hence, $(M_N, \mathcal{V}) \models \forall x(\psi')^t$, as desired. Finally, suppose that $\psi = \exists x \psi'$, $(N,V) \models \psi$, and $\mathcal{V}$ is $L$-compatible with $V$. We want to show that $(M_N, \mathcal{V}) \models \psi^t$. Since $(N,V) \models \exists x \psi'$, there must exist some valuation $V' \sim_x V$ such that $(N, V') \models \psi'$. By the induction hypothesis, for all $\mathcal{V}''$ $L$-compatible with $V'$, we have $(M_N, \mathcal{V}'') \models (\psi')^t$. Choose some formula $\varphi' \in L^{-1}(V'(x))$ (such a $\varphi'$ exists since $L$ is a surjection). Define $\mathcal{V}'$ by taking $\mathcal{V}'(y) = \mathcal{V}(y)$ for $y \neq x$ and $\mathcal{V}'(x) = \varphi'$. Clearly $\mathcal{V}'$ is $L$-compatible with $V'$. Thus, $(M_N, \mathcal{V}') \models (\psi')^t$, by the induction hypothesis. Hence, $(M_N, \mathcal{V}) \models \exists x(\psi')^t$. This completes the induction proof. ∎

We have now proved one direction of (3): for all $C \subseteq \{r,t,e\}$, if $\varphi$ is satisfiable in some $R$-model, then $\varphi^t \wedge \sigma$ is satisfiable in some structure in $\mathcal{M}_n^C(\Phi, \mathcal{X})$. For the converse, suppose that $\varphi^t \wedge \sigma$ is satisfiable in some structure $M = (S, \mathcal{K}, \mathcal{A}, \pi)$ in $\mathcal{M}_1^C(\Phi, \mathcal{X})$. If $(M,s) \models \varphi^t \wedge \sigma$, then define an $R$-model $N_{M,s}$ whose domain $D_{M,s} = \mathcal{L}_1^{K,X,A}(\Phi)$ and $R^{M,s}$ (the interpretation of $R$ in $N_{M,s}$) is $\{(\psi, \psi') : \pi(t,r) = \textbf{true}$ for all $t$ such that $(s,t) \in \mathcal{K}$, $\psi \wedge q_1 \in \mathcal{A}(t)$, and $\psi' \wedge q_2 \in \mathcal{A}(t)\}$. Note that because the domain on $N_{M,s}$ is $\mathcal{L}_1^{K,X,A}(\Phi)$, a syntactic valuation is also a valuation on $N_{M,s}$. The other direction of (3) follows immediately from the following result.

**Lemma A.8:** *For all formulas $\psi$ in negation normal form and all syntactic valuations $\mathcal{V}$, if $(M, s, \mathcal{V}) \models \psi^t \wedge \sigma$ then $(N_{M,s}, \mathcal{V}) \models \psi$.*

**Proof:** We prove the lemma by induction on the length of $\psi$. If $\psi = R(x,y)$ and $(M, s, \mathcal{V}) \models \psi^t \wedge \sigma$, then that there exists $t$ such that $(s,t) \in \mathcal{K}$, $\pi(t,r) = \textbf{true}$, $\mathcal{V}(x) \wedge q_1 \in \mathcal{A}(t)$, and $\mathcal{V}(y) \wedge q_2 \in \mathcal{A}(t)$. Since $\sigma$ implies that for all $t'$ such that $(s,t') \in \mathcal{K}$, $\mathcal{V}(x) \wedge q_1 \in \mathcal{A}(t')$ and $\mathcal{V}(y) \wedge q_2 \in \mathcal{A}(t')$, it must be the case that $\pi(t', r) = \textbf{true}$. Thus, by definition of $R^{M,s}$, it follows that $(\mathcal{V}(x), \mathcal{V}(y)) \in R^{M,s}$. Therefore, $(N_{M,s}, \mathcal{V}) \models \psi$. A similar argument applies if $\psi$ is of the form $\neg R(x,y)$. If $\psi = \psi_1 \wedge \psi_2$ or $\psi = \psi_1 \vee \psi_2$, the result follows easily from the induction hypothesis. If $\psi = \forall x \psi'$ and $(M, s, \mathcal{V}) \models \psi^t \wedge \sigma$, then, since $\sigma$ is a sentence, for all $\mathcal{V}' \sim_x \mathcal{V}$, we have $(M, s, \mathcal{V}') \models (\psi')^t \wedge \sigma$. By the induction hypothesis, it follows that $(N_{M,s}, \mathcal{V}') \models \psi'$ for all $\mathcal{V}'$ such that $\mathcal{V}' \sim_x \mathcal{V}$. Since $D_n = \mathcal{L}_1^{K,X,A}(\Phi)$, it follows that $(N_{M,s}, \mathcal{V}) \models \forall x \psi'$. Finally, suppose that $\psi = \exists x \psi'$ and $(M, s, \mathcal{V}) \models \psi^t \wedge \sigma$. Again, since $\sigma$ is a sentence, there exists some $\mathcal{V}' \sim_x \mathcal{V}$ such that $(M, s, \mathcal{V}') \models (\psi')^t \wedge \sigma$. By the induction hypothesis, it follows that $(N_{M,s}, \mathcal{V}') \models \psi'$. Thus, $(N_{M,s}, \mathcal{V}) \models \exists x \psi'$, as desired. ∎

This completes the proof of (3) and Theorem 5.1. Note that exactly the same proof works if we take $q_2 = \neg q_1$ and $r = q_1$. Therefore, the assumption that $\Phi$ includes $q_1$, $q_2$ and $r$ can be made without loss of generality. ∎

**Theorem 5.2:** *The problem of deciding if a formula in the language $\mathcal{L}_n^{\forall, K}$ is valid in $\mathcal{M}_n^C(\Phi)$ is r.e.-complete if $n \geq 2$ or if $e \notin C$.*

**Proof:** The fact that deciding validity is r.e. in all these cases follows immediately from Theorem 2.1.

To prove hardness, we start with the case that $e \in C$ and $n = 2$. For ease of exposition, we assume that $\Phi$ is countably infinite. We show at the end of the proof how to remove this assumption. Let an $R$-formula, an $R$-model, and an *countable $R$ model* be as defined in the proof of Theorem 5.1. Again, it suffices to reduce the satisfiability problem for $R$-formulas to the satisfiability problem for formulas in $\mathcal{L}_n^{\forall, K}(\Phi, \mathcal{X})$.

Assume that $\Phi$ contains the primitive propositions $p$, $q$, and $r$. Our goal is to write a modal formula that forces a model to have four types of states:

- States satisfying $\neg p \wedge \neg q$. Intuitively, these states will represent pairs $(d_1, d_2)$ of domain elements in an $R$-model.

- States satisfying $p \wedge \neg q$. Intuitively, these states represent the first element $d_1$ in a pair $(d_1, d_2)$.

- States satisfying $\neg p \wedge q$. Intuitively, these states represent the second element $d_2$ in a pair $(d_1, d_2)$.

- States satisfying $p \wedge q$. Intuitively, these states represent domain elements.

We want it to be the case that the states satisfying $\neg p \wedge \neg q$ form a $\mathcal{K}_1$ equivalence class; for each state satisfying $\neg p \wedge \neg q$, there is a $\mathcal{K}_2$-edge going to a state satisfying $p \wedge \neg q$ and one going to a state satisfying $\neg p \wedge q$. Intuitively, this triple of states represents a pair $(d_1, d_2)$, the first component of the pair, and the second component of the pair. Finally, from each state satisfying $p \wedge \neg q$ or $\neg p \wedge q$, there is a $\mathcal{K}_1$-edge to a state satisfying $p \wedge q$; the latter state is the one that determines the domain element. Finally, the primitive proposition $r$ is true at a state satisfying $\neg p \wedge \neg q$ iff $R(d_1, d_2)$ holds in the $R$-model. (We remark that this construction is somewhat similar in spirit to a construction used by Engelhardt, van der Meyden, and Moses [2005] to prove that, in the case of semantic valuations, the validity problem is $\Pi^2$- complete.) Figure 1 describes the desired situation:

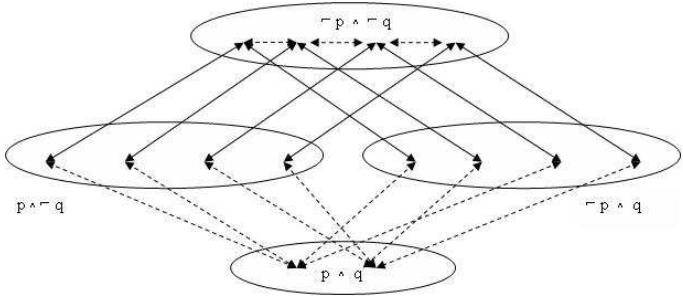

Figure 1: States.

In the figure, the $\mathcal{K}_1$ relation consists of the pairs joined by dotted lines; the $\mathcal{K}_2$ consists of the pairs liked by the continuous edges. (In both cases we omit self-loops.)

Let $atomic(x)$ be an abbreviation for the formula

$$\neg K_1 K_2 K_1 \neg (p \wedge q \wedge x) \wedge \neg \exists y (\neg K_1 K_2 K_1 \neg (x \wedge y) \wedge \neg K_1 K_2 K_1 \neg (x \wedge \neg y)).$$

Intuitively, $atomic(x)$ is true if all $K_1 K_2 K_1$-reachable worlds that satisfy $x$ agree on all sentences. We use worlds where $p \wedge q \wedge x$ holds for some atomic formula $x$ to represent elements in $d$. If two worlds satisfy the same atomic formula, then they represent the same domain element.

Let $\sigma_1$ be the modal formula that forces the set of $atomic$ formulas to be non-empty. $\sigma_1$ is an abbreviation for $\exists x (atomic(x))$.

Let $\sigma_2$ be the modal formula that, roughly speaking, forces it to be the case that if $r$ is true at some state $t$ that represents $(d_1, d_2)$ (i.e., a state where $\neg p \wedge \neg q$ is true), then $r$ is true at all states $t'$ that represent $(d_1, d_2)$. (It follows that if $\neg r$ is true at some state $t$ that represents $(d_1, d_2)$, then $\neg r$ is true at all states $t'$ that represent $(d_1, d_2)$.) The formula $\sigma_2$ is an abbreviation for

$\forall x \forall y ((atomic(x) \wedge atomic(y) \wedge$
$\neg K_1 \neg (r \wedge \neg p \wedge \neg q \wedge \neg K_2 \neg (p \wedge \neg q \wedge \neg K_1 \neg (p \wedge q \wedge x)) \wedge \neg K_2 \neg (\neg p \wedge q \wedge \neg K_1 \neg (p \wedge q \wedge y))))$
$\Rightarrow K_1 (\neg p \wedge \neg q \wedge \neg K_2 \neg (p \wedge \neg q \wedge \neg K_1 \neg (p \wedge q \wedge x)) \wedge \neg K_2 \neg (\neg p \wedge q \wedge \neg K_1 \neg (p \wedge q \wedge y)) \Rightarrow r)).$

Let $\sigma = \sigma_1 \wedge \sigma_2$.

We now translate an $R$-formula $\psi$ to an awareness formula $\psi^t$. We consider only $R$-formulas formulas in negation normal form.

- $(R(x, y))^t = atomic(x) \wedge atomic(y) \wedge \neg K_1 \neg (r \wedge \neg p \wedge \neg q \wedge \neg K_2 \neg (p \wedge \neg q \wedge \neg K_1 \neg (p \wedge q \wedge x)) \wedge \neg K_2 \neg (\neg p \wedge q \wedge \neg K_1 \neg (p \wedge q \wedge y)));$

- $(\neg R(x, y))^t = atomic(x) \wedge atomic(y) \wedge \neg K_1 \neg (\neg r \wedge \neg p \wedge \neg q \wedge \neg K_2 \neg (p \wedge \neg q \wedge \neg K_1 \neg (p \wedge q \wedge x)) \wedge \neg K_2 \neg (\neg p \wedge q \wedge \neg K_1 \neg (p \wedge q \wedge y)));$

- $(\varphi_1 \wedge \varphi_2)^t = \varphi_1^t \wedge \varphi_2^t;$

- $(\varphi_1 \vee \varphi_2)^t = \varphi_1^t \vee \varphi_2^t$;

- $(\forall x \varphi)^t = \forall x(atomic(x) \Rightarrow \varphi^t)$;

- $(\exists x \varphi)^t = \exists x(atomic(x) \wedge \varphi^t)$.

Theorem 5.2 in the case that $C \supseteq \{e\}$ follows from the following claim: If $e \in C$ and $n \geq 2$, then

For every $R$-sentence $\psi$, $\psi$ is satisfiable in an $R$-model iff $\psi^t \wedge \sigma$ is satisfiable in $\mathcal{M}_n^C(\Phi, \mathcal{X})$. (4)

To prove (4), first suppose that $\psi$ is a satisfiable $R$-sentence. As in Theorem 5.1, we can assume without loss of generality that $\psi$ is satisfied in an $R$-model $N$ with countable domain $D_N$. Let $L$ be a surjection from $\Phi - \{p, q, r\}$ to $D_N$. (Since $D_N$ is countable and $\Phi$ is countably infinite, by assumption, such a surjection exists.) Define $M_N = (S, \mathcal{K}_1, \mathcal{K}_2, \pi)$ to be the Kripke structure such that

- $S = D_N \cup (D_N \times D_N\} \cup_{i=1}^2 \{(d_1, d_2, i) : d_1, d_2 \in D_N\}$;

- $\pi(s, r) = $ **true** iff $(d_1, d_2) \in R$, and $s = (d_1, d_2)$;

- $\pi(s, p) = $ **true** iff either $s \in D_N$ or $s$ is of the form $(d_1, d_2, 1)$ for some $d_1, d_2 \in D_N$;

- $\pi(s, q) = $ **true** iff either $s \in D_N$ or $s$ is of the form $(d_1, d_2, 2)$ for some $d_1, d_2 \in D_N$;

- for all $p' \in \Phi - \{p, q, r\}$, $\pi(s, p') = $ **true** iff $L(p') = d$ and $s = d$;

- $\mathcal{K}_1(s) = (D_N \times D_N)$ for $s \in (D_N \times D_N)$, and $\mathcal{K}_1((d_1, d_2, 1)) = \mathcal{K}_1((d_2, d_1, 2)) = \mathcal{K}_1(d_1) = \{(d_1, d_2, 1), (d_2, d_1, 2), d_1\}$ for $d_1, d_2 \in D_N$;

- $\mathcal{K}_2((d_1, d_2)) = \mathcal{K}_2((d_1, d_2, 1)) = \mathcal{K}_1((d_1, d_2, 2)) = \{(d_1, d_2), (d_1, d_2, 1), (d_1, d_2, 2)\}$ for $d_1, d_2 \in D_N$, and $\mathcal{K}_2(d) = \{d\}$ for $d \in D_N$.

It is easy to check that $M_N \in \mathcal{M}_2^{r,e,t}(\Phi, \mathcal{X})$ (and hence also in $\mathcal{M}_2^C(\Phi, \mathcal{X})$ for all $C$ such that $e \in C$) and that $(M_N, (d_1, d_2)) \models \sigma$ for all $(d_1, d_2) \in D_N \times D_N$ (note that $\sigma$ is a sentence and therefore is independent of the valuation); we leave the proof to the reader. Thus, it suffices to show that there exists a state $s^* \in D_N \times D_N$ such that $(M_N, s^*) \models \psi^t$. This follows from the following result.

**Lemma A.9:** If $s^* \in D_N \times D_N$, then for every first-order sentence $\psi$ in negation normal form, if $N \models \psi$ then $(M_N, s^*) \models \psi^t$.

**Proof:** Fix $s^* \in D_N \times D_N$. We actually prove a slightly more general result. A syntactic valuation $\mathcal{V}$ is $M_N$-*compatible* with a valuation $V$ on $N$ if, for all variables $x$ and all $s \in S$, $(M_N, s, \mathcal{V}) \models x$ iff $s = V(x)$. We show that for all first-order formulas $\psi$ (not just sentences) and all valuations $V$ on $N$, if $(N, V) \models \psi$, then $(M_N, s^*, \mathcal{V}) \models \psi^t$ for all syntactic valuations $\mathcal{V}$ $M_N$-compatible with $V$. The proof is by induction on structure.

Suppose that $\psi = R(x, y)$. Then, $(N, V) \models \psi$ iff $(V(x), V(y)) \in R$. By definition of $\pi$, if $(V(x), V(y)) \in R$ then $\pi((V(x), v(y)), r) = \mathbf{true}$. Let $\mathcal{V}$ be a syntactic valuation $M_N$-compatible with $V$. By definition, $(M_N, s_1, \mathcal{V}) \models x$ iff $s_1 = V(x)$ and $(M_N, s_2, \mathcal{V}) \models y$ iff $s_2 = V(y)$. Thus, by definition of $M_N$, it is easy to see that $(M_N, s^*, \mathcal{V}) \models atomic(x) \wedge atomic(y)$. By definition of $\mathcal{K}_1$ and $\pi$, we have $(M_N, (V(x), V(y), 1), \mathcal{V}) \models \neg K_1 \neg (p \wedge q \wedge x)$ and $(M_N, (V(x), V(y), 2), \mathcal{V}) \models \neg K_1 \neg (p \wedge q \wedge y)$. By definition of $\mathcal{K}_2$ and $\pi$, it follows that $(M_N, (V(x), V(y)), \mathcal{V}) \models r \wedge \neg p \wedge \neg q \wedge \neg K_2 \neg (p \wedge \neg q \wedge \neg K_1 \neg (p \wedge q \wedge x)) \wedge \neg K_2 \neg (\neg p \wedge q \wedge \neg K_1 \neg (p \wedge q \wedge y))$. Since $(V(x), V(y)) \in \mathcal{K}_1(s^*)$, $(M_N, s^*, \mathcal{V}) \models atomic(x) \wedge atomic(y)$, and $(R(x, y))^t = atomic(x) \wedge atomic(y) \wedge \neg K_1 \neg (r \wedge \neg p \wedge \neg q \wedge \neg K_2 \neg (p \wedge \neg q \wedge \neg K_1 \neg (p \wedge q \wedge x)) \wedge \neg K_2 \neg (\neg p \wedge q \wedge \neg K_1 \neg (p \wedge q \wedge y)))$, it follows that $(M_N, s^*, \mathcal{V}) \models R(x, y)^t$ for all $\mathcal{V}$ $M_N$-compatible with $V$. A similar argument applies if $\psi$ is of the form $\neg R(x, y)$.

If $\psi = \psi_1 \wedge \psi_2$ or $\psi = \psi_1 \vee \psi_2$, the result follows easily from the induction hypothesis.

Suppose that $\psi = \exists x \psi'$, $(N, V) \models \psi$, and $\mathcal{V}$ is $M_N$-compatible with $V$. We want to show that $(M_N, s^*, \mathcal{V}) \models \psi^t$. Since $(N, V) \models \exists x \psi'$, there must exist some valuation $V' \sim_x V$ such that $(N, V') \models \psi'$. By the induction hypothesis, for all $\mathcal{V}''$ $M_N$-compatible with $V'$, we have $(M_N, s^*, \mathcal{V}'') \models (\psi')^t$. Choose some primitive proposition $p' \in L^{-1}(V'(x))$ (such a $p'$ exists since $L$ is a surjection). Define $\mathcal{V}'$ by taking $\mathcal{V}'(y) = \mathcal{V}(y)$ for $y \neq x$ and $\mathcal{V}'(x) = p'$. Clearly $\mathcal{V}'$ is $M_N$-compatible with $V'$. Thus, by the induction hypothesis, $(M_N, s^*, \mathcal{V}') \models (\psi')^t$. Since $(M_N, s^*, \mathcal{V}') \models atomic(x)$, we have $(M_N, s^*, \mathcal{V}) \models \exists x(atomic(x) \wedge (\psi')^t)$ for all $\mathcal{V}$ $M_N$-compatible with $V$.

Finally, suppose that $\psi = \forall x \psi'$, $(N, V) \models \forall x \psi'$, and $\mathcal{V}$ is $M_N$-compatible with $V$. We want to show that $(M_N, s^*, \mathcal{V}) \models \psi^t$. Since $\psi^t = \forall x(atomic(x) \Rightarrow (\psi')^t)$, we must show that $(M_N, s^*, \mathcal{V}') \models (atomic(x) \Rightarrow (\psi')^t)$ for all $\mathcal{V}' \sim_x \mathcal{V}$. Given a valuation $\mathcal{V}' \sim_x \mathcal{V}$, suppose that $(M_N, s^*, \mathcal{V}') \models atomic(x)$. It follows that there is a unique $t \in D_N$ such that $(M_N, t, \mathcal{V}') \models x$. Let $V'$ be the valuation such that $V' \sim_x V$ and $V'(x) = t$. Since $\mathcal{V}$ is $M_N$-compatible with $V$, it can be easily shown that $\mathcal{V}'$ $M_N$-compatible with $V'$. It thus follows from the induction hypothesis that $(M_N, s^*, \mathcal{V}') \models (\psi')^t$. Hence, $(M_N, s^*, \mathcal{V}) \models \forall x(atomic(x) \Rightarrow (\psi')^t)$, as desired. This completes the induction proof. ∎

To prove the other direction of (4), suppose that $\varphi^t \wedge \sigma$ is satisfiable in some structure $M = (S, \mathcal{K}_1, \mathcal{K}_2, \pi) \in \mathcal{M}_2^e(\Phi, \mathcal{X})$. If $(M, s, \mathcal{V}) \models \varphi^t \wedge \sigma$, then define an $R$-model $N_{M,s}$ whose domain $D_{M,s} = \{\varphi \in \mathcal{L}_2^K(\Phi) : (M, s) \models atomic(\varphi)\}$ and $R^{M,s}$ (the interpretation of $R$ in $N_{M,s}$) is $\{(\psi, \psi') : \pi(t, r) = \mathbf{true}$ for all $t$ such that $(s, t) \in \mathcal{K}, (M, t) \models \neg p \wedge \neg q \wedge \neg K_2 \neg (p \wedge \neg q \wedge \neg K_1 \neg (p \wedge q \wedge \psi)) \wedge \neg K_2 \neg (\neg p \wedge q \wedge \neg K_1 \neg (p \wedge q \wedge \psi'))$. Define $V$ to be $D_{M,s}$-compatible with $\mathcal{V}$ if $\mathcal{V}(x) \in D_{M,s}$ implies that $V(x) = \mathcal{V}(x)$. The other direction of (4) follows immediately from the following result.

**Lemma A.10:** *For all formulas $\psi$ in negation normal form and all syntactic valuations $\mathcal{V}$, if $(M, s, \mathcal{V}) \models \psi^t \wedge \sigma$ then $(N_{M,s}, V) \models \psi$ for all $V$ $D_{M,s}$-compatible with $\mathcal{V}$.*

**Proof:** We prove the lemma by induction on the structure of $\psi$. If $\psi = R(x,y)$ and $(M, s, \mathcal{V}) \models \psi^t \wedge \sigma$, then $(M, s, \mathcal{V}) \models atomic(x) \wedge atomic(y)$ and there exists $t$ such that $(s,t) \in \mathcal{K}_1$ and $(M, t, \mathcal{V}) \models r \wedge \neg p \wedge \neg q \wedge \neg K_2 \neg (p \wedge \neg q \wedge \neg K_1 \neg (p \wedge q \wedge x)) \wedge \neg K_2 \neg (\neg p \wedge q \wedge \neg K_1 \neg (p \wedge q \wedge y))$. Since $\sigma$ implies that for all $t'$ such that $(s,t') \in \mathcal{K}_1$ and $(M, t', \mathcal{V}) \models \neg p \wedge \neg q \wedge \neg K_2 \neg (p \wedge \neg q \wedge \neg K_1 \neg (p \wedge q \wedge x)) \wedge \neg K_2 \neg (\neg p \wedge q \wedge \neg K_1 \neg (p \wedge q \wedge y))$, it must be the case that $\pi(t', r) = \mathbf{true}$. Thus, by definition of $R^{M,s}$, it follows that $(\mathcal{V}(x), \mathcal{V}(y)) \in R^{M,s}$. Since $\mathcal{V}(x), \mathcal{V}(y) \in D_{M,s}$, $(V(x), V(y)) \in R^{M,s}$ for all $V$ $D_{M,s}$-compatible with $\mathcal{V}$. Therefore, $(N_{M,s}, V) \models \psi$ for all $V$ $D_{M,s}$-compatible with $\mathcal{V}$. A similar argument applies if $\psi$ is of the form $\neg R(x,y)$. If $\psi = \psi_1 \wedge \psi_2$ or $\psi = \psi_1 \vee \psi_2$, the result follows easily from the induction hypothesis.

Suppose that $\psi = \forall x \psi'$ and $(M, s, \mathcal{V}) \models \sigma \wedge \psi^t$. Since $\psi^t = \forall x (atomic(x) \Rightarrow (\psi')^t)$ and $\sigma$ is a sentence, for all $\mathcal{V}' \sim_x \mathcal{V}$, we have $(M, s, \mathcal{V}') \models \sigma \wedge (atomic(x) \Rightarrow (\psi')^t)$. In particular, for all $\mathcal{V}' \sim_x \mathcal{V}$ such that $(M, s, \mathcal{V}') \models atomic(x)$, we have $(M, s, \mathcal{V}') \models (\psi')^t$. By the induction hypothesis, it follows that $(N_{M,s}, V') \models \psi'$ for all $V'$ $D_{M,s}$-compatible with $\mathcal{V}'$. Suppose that $V$ is $D_{M,s}$-compatible with $\mathcal{V}$. We want to show that $(N_{M,s}, V) \models \forall x \psi'$. Consider any $V' \sim_x V$. Let $\mathcal{V}''$ be the syntactic valuation such that $\mathcal{V}'' \sim_x \mathcal{V}$ and $\mathcal{V}''(x) = V'(x)$. Clearly $V'$ is $D_{M,s}$-compatible with $\mathcal{V}''$. Since $\mathcal{V}'' \sim_x \mathcal{V}$, $(M, s, \mathcal{V}'') \models atomic(x)$, and $V'$ is $D_{M,s}$-compatible with $\mathcal{V}''$, the induction hypothesis implies that $(N_{M,s}, V') \models \psi'$. It follows that $(N_{M,s}, V') \models \psi'$. Therefore, $(N_{M,s}, V) \models \forall x \psi'$, as desired.

Finally, suppose that $\psi = \exists x \psi'$ and $(M, s, \mathcal{V}) \models \psi^t \wedge \sigma$. Since $\psi^t = \exists x (atomic(x) \wedge (\psi')^t)$ and $\sigma$ is a sentence, there exists some $\mathcal{V}' \sim_x \mathcal{V}$ such that $(M, s, \mathcal{V}') \models \sigma \wedge atomic(x) \wedge (\psi')^t$. By the induction hypothesis, it follows that $(N_{M,s}, V') \models \psi'$ for all $V'$ $D_{M,s}$-compatible with $\mathcal{V}'$. Let $V$ be $D_{M,s}$-compatible with $\mathcal{V}$. Let $V''$ be the valuation such that $V'' \sim_x V$ and $V''(x) = \mathcal{V}'(x)$. Clearly $V''$ is $D_{M,s}$-compatible with $\mathcal{V}'$. Thus, by the induction hypothesis, $(N_{M,s}, V'') \models \psi'$. Since $V'' \sim_x V$, it follows that $(N_{M,s}, V) \models \exists x \psi'$, as desired. ∎

This completes the proof of Theorem 5.2 in the case that $e \in C$. We next briefly describe the changes necessary to deal with the case that $e \notin C$.

Let $atomic'(x)$ (resp., $\sigma_1'$, $\sigma_2'$, $\sigma'$, and $\psi^T$) be the result of replacing every occurrence of $K_2$ in $atomic(x)$ (resp., $\sigma_1$, $\sigma_2$, $\sigma$, and $\psi^t$) by $K_1$. (Of course, if $t \in C$, then the $K_1 K_1 K_1$ in $atomic(x)$ can be simplified to $K_1$.) We now show that (4) holds if $C \subseteq \{r, t\}$ and $n \geq 1$. For the forward direction, if $\psi$ is satisfiable in an $R$-model $N$ with a countable domain $D_N$, we construct a structure $M_N' = (S, \mathcal{K}_1', \pi)$, where $S$ and $\pi$ are just as in the construction of $M_N$ and $\mathcal{K}_1'$ is given by

- $\mathcal{K}_1'((d_1, d_2)) = D_N \times D_N \cup \{(d_1, d_2, 1), (d_1, d_2, 2), d_1, d_2\}, \mathcal{K}_1'((d_1, d_2, 1)) = \{(d_1, d_2, 1), d_1\},$ $\mathcal{K}_1'((d_1, d_2, 2)) = \{(d_1, d_2, 2), d_2\}, \mathcal{K}_1'(d_1) = \{d_1\}$ for $d_1, d_2 \in D_N$;

It is easy to check that $M_N' \in \mathcal{M}_1^{r,t}(\Phi, \mathcal{X})$ (and hence also in $\mathcal{M}_1^C(\Phi, \mathcal{X})$ for all $C$ such that $e \notin C$) and that $(M_N', (d_1, d_2)) \models \sigma$ for all $(d_1, d_2) \in D_N \times D_N$; we leave the proof to the reader. We also leave it to the reader to check that the analogue of Lemma A.9 holds. For the converse, if $\psi^T \wedge \sigma'$ is satisfiable in a Kripke structure $M \in \mathcal{M}_1$, we construct an $R$-model satisfying $\psi$ using essentially the same construction as above, except that in defining the interpretation of $R$, we replace every occurrence of $K_2$ by $K_1$. We leave it to the reader to show that the analogue of Lemma A.10 holds.

Up to now we have assumed that $\Phi$ is infinite. However, we can apply the techniques of [Halpern 1995] to show that undecidability holds in all cases even if $|\Phi| = 1$. Suppose that $p^* \in \Phi$. We briefly sketch the argument in the case that $e \in C$. Let $q_j$ be the formula $\neg K_2 K_1 \neg(\neg p^* \wedge \neg (K_2 K_1)^j p^*)$, where $(K_2 K_1)^j$ is an abbreviation for $j$ repetitions of $K_2 K_1$. Intuitively, $q_j$ is true at a state if there is a path that leads to $p^*$ in one $K_1 K_2$-step and leads to $\neg p^*$ in an other $j$-$K_1 K_2$ steps. Let $r_1$ be $q_1$ and let $r_{j+1}$ be $q_{j+1} \wedge \neg r_1 \wedge \ldots \neg r_j$. Clearly the formulas $r_j$ are clearly mutually exclusive. In $\sigma_2$, we replace $\neg p \wedge q$ by $r_1$, replace $p \wedge \neg q$ by $r_2$, replace $r \wedge \neg p \wedge \neg q$ by $r_3$, and replace $\neg r \wedge \neg p \wedge \neg q$ by $r_4$ (so that $\neg p \wedge \neg q$ is replaced by $r_3 \vee r_4$). In $\sigma_1$, we replace $p \wedge q$ by $\neg(r_1 \vee r_2 \vee r_3 \vee r_4)$. The translation is the same, except that now the translation for $R(x, y)$ uses $r_3$ instead of $r \wedge \neg q \wedge \neg q$, and the translation for $\neg R(x, y)$ uses $r_4$. With these changes, the proof of the analogue of Lemma A.10 follows with essentially no change. To prove the analogue of Lemma A.9, we need to construct the analogue of the Kripke structure $M_N$. The construction is essentially the same as that given above, except we need to add extra states to ensure that the appropriate formulas $r_j$ holds. For example, we want to make sure that either $r_3$ or $r_4$ holds at all states in $D_N \times D_N$, so we need to add extra states to ensure that from each state in $D_N \times D_N$ the appropriate path exists. At each state in $D_N$ we ensure that $r_j$ holds for some $j \geq 5$ and that $r_j$ holds at some state in $D_N$ for each $j \geq 5$. We then replace the surjection $L$ from $\Phi - \{p, q, r\}$ to $D_N$ by a surjection from $\{r_5, r_6, \ldots\}$ to $D_N$. We leave details to the reader.

The argument in the case that $e \notin C$ proceeds along similar lines. If $t \notin C$, we use the same formulas as above, but replace $K_2 K_1$ by $K$. If $t \in C$, a slightly different set of formulas must be used; see [Halpern 1995] for details. ∎

**Theorem 5.3:** *The validity problem for the language $\mathcal{L}_1^{\forall, K}(\Phi, \mathcal{X})$ with respect to the structures in $\mathcal{M}_1^C(\Phi, \mathcal{X})$ for $C \supseteq \{e\}$ is decidable.*

**Proof:** First, consider the case $C = \{r, e, t\}$. We use ideas originally due to Fine [1969]. The technical details follow closely the decidability proof given by Engelhardt, van der Meyden, and Su [2003] for the case where the semantics is given using semantic valuations rather than syntactic valuations. The proof proceeds by an elimination of quantifiers. Following Fine [1969], we say that a world $w$ in a structure $M$ is *describable by a sentence* $\varphi$ is $(M, w) \models \varphi$ and for all worlds $w'$, if $(M, w) \models \varphi$, then for all sentences $\psi$, $(M, w) \models \psi$ iff $(M, w') \models \psi$. A world is *describable* if it is describable by some formula $\varphi$.

Let $describable(\varphi)$ be an abbreviation for

$$\neg K \neg \varphi \wedge \neg \exists y (\neg K \neg (\varphi \wedge y) \wedge \neg K \neg (\varphi \wedge \neg y)).$$

Intuitively, $describable(\varphi)$ is satisfiable in a structure iff there is a world in the structure describable by $\varphi$. Let $C_k \varphi$ be the formula that is satisfiable in a structure $M$ iff there are at least $k$ distinct describable worlds where $\varphi$ is true that the agent considers possible, where two worlds are distinct if they disagree on the truth value of at least one formula. That is, $C_k \varphi$ for $k \geq 1$ is

an abbreviation for

$$\exists x_1 \ldots \exists x_k(\wedge_{1 \leq i < j \leq k} \neg K(x_i \Leftrightarrow x_j) \wedge \wedge_{i=1}^k (describable(x_i) \wedge \neg K \neg (x_i \wedge \varphi)))$$

Let $E_k\varphi$ be an abbreviation for $C_k\varphi \wedge \neg C_{k+1}\varphi$. Note that $E_k\varphi$ is satisfied in a structure $M$ where the $\mathcal{K}$ relation is universal iff there are exactly $k$ distinct describable worlds where $\varphi$ is true.

With semantic valuations, it is not hard to show that $\neg K \neg \varphi \Leftrightarrow C_1 \varphi$ is valid. But this is not the case if we use syntactic valuations. For example, let $M = (W, \mathcal{K}, \pi)$ be the structure where $\mathcal{K}$ is universal and for each of the (uncountably many) truth assignments $v$ to the countably infinite set of primitive propositions in $\Phi$, there is a unique world $w_v$ where $\pi(w_v) = v$. Each of these worlds is clearly distinct. Since there are only countably many formulas and uncountably many worlds, there must be uncountably many undescribable worlds in this structure. (In fact, a symmetry argument shows that in this structure no world is describable.) Thus, we need to distinguish structures where $\varphi$ is satisfiable but none of the worlds in which $\varphi$ is satisfiable is describable, and structures where $\varphi$ is not satisfiable. (Both types of structures satisfy $\neg C_1 \varphi$.) Let $E_\infty \varphi$ be an abbreviation for $\neg K \neg \varphi \wedge \neg C_1 \varphi$ and let $E_0 \varphi$ be an abbreviation for $K \neg \varphi$. (Note that $K \neg \varphi \Rightarrow \neg C_1 \varphi$ is valid.) It is not hard to show that if $E_\infty \varphi$ is satisfied in a structure $M$, then there are actually infinitely many distinct worlds at which $\varphi$ is true (although none of them is describable). Finally, if $l \neq \infty$, let $M_{l,N}\varphi$ be an abbreviation for $E_l \varphi$ if $l < N$ and for $C_l\varphi$ if $l \geq N$.

Let $\mathbf{p} = (p_1, \ldots, p_m)$ be a vector of primitive propositions and propositional variables. Define a *point atom* for $\mathbf{p}$ to be a formula of the form $l_1 \wedge \ldots l_m$ where each $l_i$ is either $p_i$ or $\neg p_i$. Let $PA(\mathbf{p})$ denote the set of point atoms of $\mathbf{p}$. Given a point atom $a$ for $\mathbf{p}$ and a number $N$, define an $N$-*bounded count* of $a$ to be a formula of the form $E_l a$ where $l < N$ or $l = \infty$, or $C_N a$. Define a $(\mathbf{p}, k)$-atom to be a formula of the form

$$a \wedge \wedge_{b \in PA(\mathbf{p})} c_b,$$

where $a$ is a point atom for $\mathbf{p}$ and $c_b$ is an $2^k$-bounded count of $b$ for each $b \in PA(\mathbf{p})$, such that $c_a$ is not $E_0 a$. We write At($\mathbf{p}$,k) for the set of $(\mathbf{p}, k)$-atoms. These atoms have the following properties.

**Lemma A.11:**

(a) *If $A, A' \in At(\mathbf{p}, k)$ are distinct atoms, then $\mathcal{M}_1^{r,e,t} \models \neg(A \wedge A')$.*

(b) *$\mathcal{M}_1^{r,e,t} \models \vee_{A \in At(\mathbf{p},k)} A$.*

(c) *If $A, A' \in At(\mathbf{p}, k)$, then either $\mathcal{M}_1^{r,e,t} \models A \Rightarrow \neg K \neg A'$ or $\mathcal{M}_1^{r,e,t} \models A \Rightarrow K \neg A'$. Moreover, we can effectively decide which holds.*

(d) *If $A \in At(\mathbf{p}, k+1)$ and $B \in At(\mathbf{p} \cdot x, k)$ where $x$ does not occur in $\mathbf{p}$, then either $\mathcal{M}_1^{r,e,t} \models A \Rightarrow \exists x B$ or $\mathcal{M}_1^{r,e,t} \models A \Rightarrow \neg \exists x B$. Moreover, we can effectively decide which holds.*

**Proof:** For part (a), note that if $A$ and $A'$ are distinct $(\mathbf{p},k)$-atoms, then either they differ in their point atom or they differ in the counting of some point atom. Therefore, it easily follows that $\mathcal{M}_1^{r,e,t} \models \neg(A \wedge A')$. For part (b), note that in each world in a structure $M \in \mathcal{M}_1^{r,e,t}$, exactly one point atom is true, and for each point atom, exactly one $2^k$-bounded count holds. For part (c), it can be easily checked that $\mathcal{M}_1^{r,e,t} \models A \Rightarrow \neg K \neg A'$ iff $A$ and $A'$ agree on all the conjuncts that are $2^k$-bounded counts of some atom in $PA(\mathbf{p})$ and that if $a'$ is the point atom in $A'$, $2^k$-bounded count of $a'$ (in both $A$ and $A'$) is not $E_0 a'$. Otherwise, it is easy to check that $\mathcal{M}_1^{r,e,t} \models A \Rightarrow K \neg A'$.

For part (d), suppose that $A$ is a $(\mathbf{p}, k+1)$-atom and $B$ is a $(\mathbf{p} \cdot x, k)$-atom. Define an *x-partition* of a $2^{k+1}$-bounded count $c_b$ of an atom $b \in PA(\mathbf{p})$ to be a formula of the form $e^+ \wedge e^-$, where $e^+$ is a $2^k$-bounded count of $b \wedge x$, $e^-$ is a $2^k$-bounded count of $b \wedge \neg x$, and the following constraints are satisfied:

1. if $c_b = E_l b$ for $l \neq \infty$, then $e^+ = M_{l_+, 2^k}(b \wedge x)$ and $e^- = M_{l_-, 2^k}(b \wedge \neg x)$ where $l_+ + l_- = l$;

2. if $c_b = C_{2^{k+1}} b$, then

    (a) $e^+ = C_{2^k}(b \wedge x)$ and $e^- = C_{2^k}(b \wedge \neg x)$, or
    
    (b) $e^+ = C_{2^k}(b \wedge x)$ and $e^- = E_{l_-}(b \wedge \neg x)$ where $l_- < 2^k$, or
    
    (c) $e^+ = E_{l_+}(b \wedge x)$ and $e^- = C_{2^k}(b \wedge \neg x)$ where $l_+ < 2^k$;

3. if $c_b = E_\infty b$, then $e^+ = E_\infty(b \wedge x)$ and $e^- = E_\infty(b \wedge \neg x)$.

We claim that if $e^+ \wedge e^-$ is an $x$-partition of $c_b$, then $\mathcal{M}_1^{r,e,t} \models c_b \Rightarrow \exists x(e^+ \wedge e^-)$. To see that, first suppose that $c_b = E_l b$ for $l \neq \infty$. Then we can suppose without loss of generality that $M \models c_b$, so that $b$ is satisfiable in exactly $l$ describable worlds. Let $\varphi_1, \ldots, \varphi_l$ be the formulas describing these worlds, and let $\varphi = \vee_{i=1}^{l_+} \varphi_i$. It is easy to see that $\mathcal{M}_1^{r,e,t} \models c_b \Rightarrow (e^+ \wedge e^-)[x/\varphi]$, as desired. Essentially the same argument works if $c_b = C_{2^{k+1}} b$ except that we replace $l$ by $2^{k+1}$ and, in addition, we replace $l_+$ in the definition of $\varphi$ by $2^k$ in case (a); replace $l_+$ by $l_-$ and substituting $x$ for $\neg \varphi$ instead of $\varphi$ in case (b). (No further changes are needed in case (c).) If $c_b = E_\infty b$, then $b$ is satisfied at infinitely many distinct worlds, none of which are describable. Thus, there must exist some formula $x$ such that both $b \wedge x$ and $b \wedge \neg x$ are satisfiable. Moreover, each of them must be satisfied in infinitely many distinct worlds, none of which are describable. For if, say, $b \wedge x$ were satisfied in only finitely many distinct worlds, it is easy to show that each of these worlds are describable, from which it follows that $b$ is satisfied in some describable world.

A similar argument shows that if $e^+ \neq E_0(b \wedge x)$, then $\mathcal{M}_1^{r,e,t} \models b \wedge c_b \Rightarrow \exists x(b \wedge x \wedge e^+ \wedge e^-)$, and if $e^- \neq E_0(b \wedge \neg x)$, then $\mathcal{M}_1^{r,e,t} \models b \wedge c_b \Rightarrow \exists x(b \wedge \neg x \wedge e^+ \wedge e^-)$. Conversely, a simple counting argument shows that if $e^+$ is a $2^k$-bounded count of $b \wedge x$, $e^-$ is a $2^k$-bounded count of $b \wedge \neg x$, and $e^+ \wedge e^-$ is not an $x$ partition of $c_b$, then $\mathcal{M}_1^{r,e,t} \models c_b \Rightarrow \neg \exists x(e^+ \wedge e^-)$.

We can assume that $A$ has the form $a \wedge \wedge_{b \in PA(\mathbf{p})} c_b$, while $B$ has the form $a' \wedge \wedge_{b \in PA(\mathbf{p})}(c_b^+ \wedge c_b^-)$, where $c_b^+$ is a $2^k$-bounded count of $b \wedge x$ and $c_b^-$ is a $2^k$-bounded count of $b \wedge \neg x$. We say that $B$ is $x$-*compatible with* $A$ if either $a' = a \wedge x$ and $c_a^+ \neq E_0(a \wedge x)$, or $a' = a \wedge \neg x$ and $c_a^- \neq E_0(a \wedge \neg x)$, and moreover, for all point atoms $b \in PA(\mathbf{p})$, we have that $c_b^+ \wedge c_b^-$ is an $x$-partition of $c_b$. Thus, if $B$ is $x$-compatible with $A$, it follows from the observations of the previous paragraph that

$$\mathcal{M}_1^{r,e,t} \models A \Rightarrow \exists x(a' \wedge c_a^+ \wedge c_a^-) \wedge \wedge_{b \in (PA(\mathbf{p}-\{a\}))} \exists x(c_b^+ \wedge c_b^-).$$

We now show that

$$\mathcal{M}_1^{r,e,t} \models (\exists x(a' \wedge c_a^+ \wedge c_a^-) \wedge \wedge_{b \in (PA(\mathbf{p})-\{a\})} \exists x(c_b^+ \wedge c_b^-)) \Rightarrow \exists x B.$$

Suppose that $(M, w, \mathcal{V}) \models \exists x(a' \wedge c_a^+ \wedge c_a^-) \wedge \wedge_{b \in (PA(\mathbf{p})-\{a\})} \exists x(c_b^+ \wedge c_b^-)$ for some $M \in \mathcal{M}_1^{r,e,t}$. Then $(M, w, \mathcal{V}) \models (a' \wedge c_a^+ \wedge c_a^-)[x/\varphi_a]$ for some formula quantifier-free sentence $\varphi_a$. Similarly, for every $b \in PA(\mathbf{p}) - \{a\}$, there exists a quantifier-free sentence $\varphi_b$ such that $(M, w, \mathcal{V}) \models (c_b^+ \wedge c_b^-)[x/\varphi_b]$. Note that, for each point atom $c \in PA(\mathbf{p})$, we can replace the formulas $\varphi_c$ with any formula $\psi_c$ that agrees with $\varphi_c$ on $c$. That is, if $M \models (a \wedge \varphi_a) \Leftrightarrow (a \wedge \psi_a)$, then $(M, w.\mathcal{V}) \models (a' \wedge c_a^+ \wedge c_a^-)[x/\psi_a]$; similarly, if $M \models (b \wedge \varphi_b) \Leftrightarrow (b \wedge \psi_b)$, then $(M, w, \mathcal{V}) \models c_b^+ \wedge c_b^-)[x/\psi_b]$. Let $\psi = \vee_{c \in PA(\mathbf{p})}(c \wedge \varphi_c)$. It is easy to see that $\psi$ agrees with each of the formulas $\varphi_c$ on $c$. It follows that

$$(M, w, \mathcal{V}) \models (a' \wedge c_a^+ \wedge c_a^-)[x/\psi] \wedge \wedge_{b \in (PA(\mathbf{p})-\{a\})}(c_b^+ \wedge c_b^-)[x/\psi].$$

Note that $\psi$ may mention variables, since the point atoms in $PA(\mathbf{p})$ may mention variables. (Recall that $\{\mathbf{p}\}$ may include propositional variables.) However, let $\psi'$ be the sentence that results by replacing each variable $y$ in $\psi$ by $\mathcal{V}(y)$. Clearly $\psi'$ is a quantifier-free sentence, and it is easy to see that

$$(M, w, \mathcal{V}) \models (a' \wedge c_a^+ \wedge c_a^-)[x/\psi'] \wedge \wedge_{b \in (PA(\mathbf{p})-\{a\})}(c_b^+ \wedge c_b^-)[x/\psi'].$$

Thus, $(M, w) \models \exists x B$, as desired. It follows that if $B$ is $x$-compatible with $A$, then $\mathcal{M}_1^{r,e,t} \models A \Rightarrow \exists x B$.

On the other hand, suppose that $B$ is not $x$-compatible with $A$. Then we have that either (1) $a'$ is not of the form $a \wedge x$ or $a \wedge \neg x$; (2) $a'$ is of the form $a \wedge x$ but $c_a^+ = E_0(a \wedge x)$; (3) $a'$ is of the form $a \wedge \neg x$ but $c_a^- = E_0(a \wedge \neg x)$; or (4) there exists a point atom $b$ such that $c_b^+ \wedge c_b^-$ is not an $x$-partition of $c_b$. In each case, it is immediate that $\mathcal{M}_1^{r,e,t} \models A \Rightarrow \neg \exists x B$. ∎

Let $\mathcal{L}_1^{\forall,K}(\mathbf{p}, k)$ consist of all formulas in $\mathcal{L}_1^{\forall,K}(\mathbf{p})$ with depth of quantification at most $k$. For a formula $\psi \in \mathcal{L}_1^{\forall,K}(\mathbf{p}, k)$, let $At(\mathbf{p}, k, \psi) = \{A \in At(\mathbf{p}, k) : \mathcal{M}_1^{r,e,t} \models A \Rightarrow \psi\}$.

**Lemma A.12:** *For all $\varphi \in \mathcal{L}_1^{\forall,K}(\mathbf{p}, k)$, $At(\mathbf{p}, k) = At(\mathbf{p}, k, \varphi) \cup At(\mathbf{p}, k, \neg\varphi)$. Moreover, the sets $At(\mathbf{p}, k, \varphi)$ and $At(\mathbf{p}, k, \neg\varphi)$ are effectively computable.*

**Proof:** We proceed by induction on $k$ with a subinduction on the structure of $\varphi$. For all $k$, the statement is immediate if $\varphi$ is a primitive proposition or propositional variable in $\{\mathbf{p}\}$, and the result follows easily from the induction hypothesis if $\varphi$ is of the form $\neg\varphi'$ or of the form $\varphi_1 \wedge \varphi_2$. If $\varphi$ is of the form $K\varphi'$, by Lemma A.11(c), we can effectively compute the set $At(\mathbf{p}, k, K\neg B)$ for each $(\mathbf{p}, k)$-atom $B$. It is easy to see that $\mathcal{M}_1^{r,e,t} \models A \Rightarrow K\varphi$ iff $\mathcal{M}_1^{r,e,t} \models A \Rightarrow K\neg B$ for all $B \in At(\mathbf{p}, k, \neg\varphi')$. Thus, $At(\mathbf{p}, k, K\varphi') = \cap_{B \in At(\mathbf{p},k,\neg\varphi')} At(\mathbf{p}, k, K\neg B)$. Moreover, if $\mathcal{M}_1^{r,e,t} \models A \Rightarrow \neg K\neg B$ for some $B \in At(\mathbf{p}, k, \neg\varphi')$, then $\mathcal{M}_1^{r,e,t} \models A \Rightarrow \neg K\varphi'$. Thus, $At(\mathbf{p}, k, \neg K\varphi') \supseteq \cup_{B \in At(\mathbf{p},k,\neg\varphi')} At(\mathbf{p}, k, \neg K\neg B)$. It follows from Lemma A.11(c) that $\cup_{B \in At(\mathbf{p},k,\neg\varphi')} At(\mathbf{p}, k, \neg K\neg B) = \overline{\cap_{B \in At(\mathbf{p},k,\neg\varphi')} At(\mathbf{p}, k, K\neg B)} = \overline{At(\mathbf{p}, k, K\varphi')}$. Since $At(\mathbf{p}, k, \neg K\varphi')$ and $At(\mathbf{p}, k, K\varphi')$ are clearly disjoint, it follows that

$$At(\mathbf{p}, k, \neg K\varphi') = \cup_{B \in At(\mathbf{p},k,\neg\varphi')} At(\mathbf{p}, k, \neg K\neg B),$$

and that both $At(\mathbf{p}, k, K\varphi')$ and $At(\mathbf{p}, k, \neg K\varphi')$ are effectively computable. Finally, if $\varphi = \forall x\varphi'$, similar arguments using Lemma A.11(d) show that $At(\mathbf{p}, k, \forall x\varphi') = \cap_{B \in At(\mathbf{p},k-1,\neg\varphi')} At(\mathbf{p}, k-1, \forall\neg B)(= \cap_{B \in At(\mathbf{p},k-1,\neg\varphi')} At(\mathbf{p}, k-1, \neg\exists x B)$ and $At(\mathbf{p}, k, \neg\forall x\varphi') = \cup_{B \in At(\mathbf{p},k-1,\neg\varphi')} At(\mathbf{p}, k-1, \neg\forall\neg B)$. Again, by Lemma A.11(d), these sets are effectively computable. ∎

To complete the proof of Theorem 5.3 for the case $C = \{r, e, t\}$, suppose that $\varphi \in \mathcal{L}_1^{\forall,K}(\Phi, \mathcal{X})$. Then there exists some finite $\mathbf{p}$ such that $\varphi \in \mathcal{L}_1^{\forall,K}(\mathbf{p})$. We claim that $\varphi$ is valid iff $At(\mathbf{p}, k, \varphi) = At(\mathbf{p}, k)$. The fact that $\varphi$ is valid if $At(\mathbf{p}, k, \varphi) = At(\mathbf{p}, k)$ follows immediately from Lemma A.11(b). For the converse, note that if $At(\mathbf{p}, k, \varphi) \neq At(\mathbf{p}, k)$, then by Lemma A.12, $At(\mathbf{p}, k, \neg\varphi) \neq \emptyset$. It is easy to see that each atom in $At(\mathbf{p}, k)$ is satisfiable in some structure in $\mathcal{M}_1^{r,e,t}$. If $M$ is a structure in $\mathcal{M}_1^{r,e,t}$ satisfying $A \in At(\mathbf{p}, k, \neg\varphi)$, then $M$ also satisfies $\neg\varphi$, showing that $\varphi$ is not valid. Finally, by Lemma A.12, we can effectively compute $At(\mathbf{p}, k, \varphi)$ and check if $At(\mathbf{p}, k, \varphi) = At(\mathbf{p}, k)$.

Thus, we have dealt with the case that $C = \{r, e, t\}$. It is well known [Fagin, Halpern, Moses, and Vardi 1995, Lemma 3.1.5] and easy to show that a reflexive Euclidean relation is transitive. Thus, $\mathcal{M}^{\{r,e,\}} = \mathcal{M}^{\{r,e,t\}}$, so we have also dealt with the case that $C = \{r, e\}$.

For the case $C = \{e, t\}$, essentially the same proof works. We briefly list the required modifications:

- We define a formula $indist$ that is true if a world indistinguishable from the current world (in the sense that the same formulas are true in both worlds) is considered possible. $indist$ is an abbreviation for:

$$\exists x(describable(x) \wedge \forall y(y \Leftrightarrow \neg K\neg(x \wedge y))).$$

Note that $indist$ is guaranteed to hold in a world where the accessibility relation is reflexive.

- We modify the definition of $(\mathbf{p}, k)$-atom. We define a $(\mathbf{p}, k)$-atom to to include a conjunct saying whether $indist$ holds. Thus, we define a $(\mathbf{p}, k)$-atom to have one of the following forms:

    - $a \wedge indist \wedge \wedge_{b \in PA(\mathbf{p})} c_b$, where $c_a \neq E_0 a$; or

- $a \wedge \neg indist \wedge \wedge_{b \in PA(\mathbf{p})} c_b$.

Note that $a \wedge \neg indist \wedge E_0 a$ is satisfiable in $\mathcal{M}^{\{e,t\}}$, since the accessibility relation no longer needs to be reflexive, but $a \wedge indist \wedge E_0 a$ is not.

- We replace $\mathcal{M}^{r,e,t}$ by $\mathcal{M}^{e,t}$ throughout the statement and proof of Lemma A.11.

- In the proof of Lemma A.11(c), we have $\mathcal{M}_1^{e,t} \models A \Rightarrow K \neg A'$ not only in the case that $A$ and $A'$ disagree on the $2^k$-bounded count of some atom in $PA(\mathbf{p})$, but also if $\neg indist$ is one of the conjuncts of $A'$. This is true since all structures in $\mathcal{M}^e$, and hence in $\mathcal{M}^{e,t}$, the $\mathcal{K}$ relations satisfy *secondary reflexivity*: if $M = (S, \mathcal{K}, \pi) \in \mathcal{M}^e$ and $(s,t) \in \mathcal{K}$, then it is easy to check that $(t,t) \in \mathcal{K}$. Thus, $indist$ holds at $t$.

- In the proof of Lemma A.11(d), we modify the definition of $x$-compatibility. We now say that $B$ is $x$-compatible with $A$ if either

  (a) $indist$ is a conjunct of both $A$ and $B$, and all the previous conditions for $x$-compatibility hold; or

  (b) $\neg indist$ is a conjunct of both $A$ and $B$, and all the preivous conditions for $x$-compatibilty hold except that we do not require that $c_a^+ \neq E_0(a \wedge x)$ or $c_a^- \neq E_0(a \wedge \neg x)$.

  It is easy to show that $\mathcal{M}^{\{e,t\}} \models A \Rightarrow \exists x B$ if $B$ is $x$-compatible with $A$ and that $\mathcal{M}^{\{e,t\}} \models A \Rightarrow \neg \exists x B$ if $B$ is not $x$-compatible with $A$.

The argument for the case $C = \{e\}$ is similar to that for $C = \{e,t\}$. It depends on the following semantics characterization of satisfiability with respect to structures in $\mathcal{M}_1^e$, similar in spirit to corresponding characterizations for $\mathcal{M}_1^{ret}$ and $\mathcal{M}_1^{rst}$ (see [Fagin, Halpern, Moses, and Vardi 1995, Proposition 3.1.6]): A formula is satisfiable in $\mathcal{M}_1^e$ iff there exists some structure $M$ such that $(M, s_0) \models \varphi$, where $M = (\{s_0\} \cup S \cup S', \pi, \mathcal{K})$, and (a) $S$ and $S'$ are disjoint sets of states; (b) if $S = \emptyset$ then $S' = \emptyset$, (c) $\mathcal{K}(s_0) = S$; (d) $\mathcal{K}(s) = S \cup S'$ if $s \in S \cup S'$; and (e) $|\{s_0\} \cup S \cup S'| \leq |\varphi|$ [Halpern and Rêgo 2006a].

Given this characterization, it can be seen that for each point atom $b$ we must count not only the number of describable worlds where $b$ is true that an agent considers possible, but also the number of describable worlds that an agent considers possible that he considers possible. Define $indist^{KK}$, $N$-bounded $KK$-count, $C_k^{KK}\varphi$, and $E_k^{KK}\varphi$ by replacing every occurrence of $K$ by $KK$ in the definitions of $indist$, $N$-bounded count, $C_K\varphi$, and $E_k\varphi$, respectively. Since the $\mathcal{K}$ relation in structures in $\mathcal{M}_1^e$ satisfies secondary reflexivity and the Euclidean property, it is easy to check that $\mathcal{M}_1^e \models indist \Rightarrow indist^{KK}$, $\mathcal{M}_1^e \models E_\infty \varphi \Rightarrow \neg E_0^{KK} \varphi$, $\mathcal{M}_1^e \models E_k \varphi \Rightarrow C_k^{KK} \varphi$, and $\mathcal{M}_1^e \models C_N \varphi \Rightarrow C_N^{KK} \varphi$.

We now modify the definition of $(\mathbf{p}, k)$-atom to include a description of what is true at the worlds that an agent considers possible that he considers possible. Thus, we now take a $(\mathbf{p}, k)$-atom to have the form

- $a \wedge indist \wedge indist^{KK} \wedge_{b \in PA(\mathbf{p})} (c_b \wedge c_b^{KK})$,

- $a \wedge \neg indist \wedge indist^{KK} \wedge_{b \in PA(\mathbf{p})} (c_b \wedge c_b^{KK})$, or

- $a \wedge \neg indist \wedge \neg indist^{KK} \wedge_{b \in PA(\mathbf{p})} (c_b \wedge c_b^{KK})$,

where (a) $c_b$ (resp., $c_b^{KK}$) is a $2^k$-bounded count (resp., $KK$-count) for all $b \in PA(\mathbf{p})$, (b) $c_a \neq E_0 a$ if $indist$ is a conjunct, (c) $c_a^{KK} \neq E_0^{KK} a$ if $indist^{KK}$ is a conjunct, (d) if $c_b = E_l b$ and $l < 2^k$, then either $c_b^{KK} = E_m^{KK} b$ and $l \leq m < \infty$, or $c_b^{KK} = C_{2^k}^{KK} b$, (e) if $c_b = E_\infty b$, then $c_b^{KK} \neq E_0^{KK} b$, and (f) if $c_b = C_{2^k} b$, then $c_b^{KK} = C_{2^k}^{KK} b$.

The same ideas used to prove Lemma A.11 for the case of $C = \{r, e, t\}$ can now be used to prove an analogous result for the case $C = \{e\}$; we omit details here. The rest of the proof is identical to that of the case $C = \{r, e, t\}$, replacing every occurrence of $\mathcal{M}_1^{r,e,t}$ by $\mathcal{M}_1^e$. ∎